\documentclass[
 aps,
 prd,
 onecolumn,
 superscriptaddress,
 nofootinbib,
 longbibliography,
 floatfix,
 amsmath,
 amssymb,
 10pt
]{revtex4-2}

\usepackage[T1]{fontenc}
\usepackage[utf8]{inputenc}
\usepackage{newtxtext,newtxmath}
\usepackage{microtype}
\usepackage[dvipsnames]{xcolor}
\usepackage{mathtools,bm}
\usepackage{graphicx}
\usepackage{booktabs}
\usepackage{multirow}
\usepackage{orcidlink}
\usepackage{hyperref}
\usepackage{bookmark}
\usepackage{flafter}
\usepackage{placeins}
\allowdisplaybreaks
\definecolor{darkblue}{rgb}{0,0,0.5}
\hypersetup{
 pdfstartview={FitH},
 pdftitle={Thermal and optical signatures of a magnetic AdS black hole in dark matter and quintessence},
 pdfauthor={Mohsen Fathi, Marco Olivares, Carlos E. Romero-Figueroa, J. R. Villanueva},
 pdfsubject={Hawking radiation, weak and strong lensing, black hole shadows, and EHT size tests},
 pdfkeywords={magnetic black holes, dark matter, quintessence, Hawking radiation, weak lensing, strong lensing, black hole shadow},
 colorlinks=true,
 linkcolor=darkblue,
 citecolor=darkblue,
 urlcolor=darkblue
}

\graphicspath{{figures/}}
\newcommand{\dd}{\mathrm{d}}
\newcommand{\cK}{\mathcal{K}}

\newcommand{\eps}{\varepsilon}

\begin{document}

\title{Thermal and optical signatures of a magnetic AdS black hole in dark matter and quintessence}

\author{Mohsen Fathi\orcidlink{0000-0002-1602-0722}}
\email{mohsen.fathi@ucentral.cl}
\affiliation{Centro de Investigaci\'{o}n en Ciencias del Espacio y F\'{i}sica Te\'{o}rica (CICEF), Universidad Central de Chile, La Serena 1710164, Chile}
\author{Marco Olivares\orcidlink{0009-0003-4533-5875}}
\email{marco.olivaresr@mail.udp.cl}
\affiliation{Facultad de Ingenier\'{i}a y Ciencias, Universidad Diego Portales, Avenida Ej\'{e}rcito Libertador 441, Santiago, Chile}

\author{Carlos E. Romero-Figueroa\orcidlink{0000-0001-5548-7766}}
\email{carlosed.romero@correo.nucleares.unam.mx}
\affiliation{Instituto de Ciencias Nucleares, Universidad Nacional Aut\'{o}noma de M\'{e}xico (UNAM), Mexico City, Mexico}

\author{J. R. Villanueva\orcidlink{0000-0002-6726-492X}}
\email{jose.villanueva@uv.cl}
\affiliation{Instituto de F\'{i}sica y Astronom\'{i}a, Universidad de Valpara\'{i}so, Avenida Gran Breta\~na 1111, Valpara\'{i}so, Chile}

\date{\today}

\begin{abstract}
We study a static magnetic anti-de Sitter black hole surrounded by perfect-fluid dark matter and quintessence. We follow the geometry from the horizon to the observer. We first examine the horizons and curvature. We then calculate the Hawking temperature and sparsity, weak light bending, the photon sphere, the shadow, strong-deflection coefficients, relativistic images, and simple Event Horizon Telescope size intervals. Because the model has several parameters, we use a one-dimensional benchmark path labelled by $\eta$. The point $\eta=0$ is Schwarzschild, while $\eta=1$ is the selected full configuration. This path is only a clear way to present the combined changes; it is not a new coupling. We derive the main relations step by step and test the weak- and strong-field approximations with direct numerical geodesics. Along the path, the outer horizon, photon sphere, critical shadow scale, and relativistic Einstein rings move outward. A formal solar-limb comparison gives $\eta\lesssim2.06\times10^{-5}$ if the same local environment is placed around the Sun, but this is a source-dependent analogue rather than a universal bound. With the central mass and distance values used here, the M87* diameter allows $0\leq\eta\leq1$, while the adopted Sgr~A* interval gives $\eta\lesssim0.665$. Longer derivations, error propagation, and numerical checks are collected in the appendices.

\end{abstract}

\maketitle

\tableofcontents

\section{Introduction}\label{sec:intro}

Black holes connect several parts of physics. General relativity describes their geometry, horizon mechanics gives them entropy and temperature, and quantum fields predict Hawking radiation \cite{Bekenstein1973,Hawking1975,Chandrasekhar1983}. They are now also observational objects. Gravitational waves test their dynamics, while very-long-baseline interferometry probes structures close to the horizon. The Event Horizon Telescope (EHT) has produced ring-like images of M87* and Sgr~A* and has opened a new way to test compact-object geometries \cite{EHT2019I,EHT2019II,EHT2019III,EHT2019IV,EHT2019V,EHT2019VI,EHT2021VII,EHT2021VIII,EHT2022I,EHT2022II,EHT2022III,EHT2022IV,EHT2022V,EHTMetric2024}.

The black hole shadow is determined by the propagation of light in the surrounding spacetime. When rays are traced backward, some escape to the illuminating region and others are captured. In a static spherical spacetime, the boundary between these two families is fixed by the unstable circular photon orbit and its critical impact parameter. This description builds on Synge’s analysis of the escape cone and the classical theory of null geodesics \cite{Synge1966,Chandrasekhar1983}. Later studies developed the modern language of black hole images and shadows \cite{Luminet1979,FalckeMeliaAgol2000,Takahashi2004,Abdujabbarov2013,Atamurotov2013,Grenzebach2014,PerlickTsupkoBisnovatyi2015}, with many extensions to rotating metrics and model-independent tests \cite{Johannsen2016,JohannsenEtAl2016,CunhaHerdeiro2018,CunhaRodriguez2018,PerlickTsupko2022,HiokiMaeda2009,Psaltis2016}. One point should remain clear: the geometrical critical curve is not always the brightest ring in an observed image. The bright ring also depends on the emitting flow, optical depth, observing frequency, and instrumental response \cite{Gralla2019,Johnson2020,BronzwaerFalcke2021,GrallaLupsasca2020,Himwich2020,ChaelInner2021,Wielgus2021,Paugnat2022}. Recent studies have further explored connections between shadows, photon rings, phase transitions, and the microscopic properties of black holes \cite{Ladino:2024ned,LADINO2025117031,ladino2026probing,ladino2026naked}. In this context, geometric approaches to black hole thermodynamics have also been proposed to relate black hole shadows to their thermodynamic behavior \cite{quevedo2026geometrothermodynamics}. However, connecting these geometrical observables to actual astronomical images requires accounting for the physical processes that shape the observed emission.

A realistic image also needs an emission model. Thin-disk studies began with the Novikov--Thorne and Page--Thorne models and later grew into full ray-tracing and radiative-transfer calculations \cite{NovikovThorne1973,PageThorne1974,Cunningham1975,BeckwithDone2005,BromleyEtAl1997,BroderickLoeb2006,DexterAgol2009,YounsiEtAl2012,FathiCruz2023}. A real interferometric comparison further needs visibilities, closure quantities, calibration, and scattering \cite{FishEtAl2014,Chael2016,Chael2018,BlackburnEtAl2020}. We do not attempt that full problem here. Our aim is simpler: we calculate the geometrical optical scales and use the EHT angular sizes only as consistency intervals. This limitation is important because our spacetime is static, while M87* and Sgr~A* are expected to rotate \cite{Kocherlakota2021,BambiEtAl2019,VagnozziEtAl2023,GiddingsPsaltis2018,Zakharov2014}.

Beyond the characterization of black hole shadows, gravitational lensing provides another direct link between spacetime geometry and astronomical observations through the deflection of light rays. Weak lensing mainly probes the outer region. Strong lensing follows rays that pass close to the photon sphere and may wind around the black hole several times. Darwin solved the Schwarzschild bending problem, while Virbhadra, Ellis, Bozza, and others developed the theory of relativistic images \cite{Darwin1959,VirbhadraEtAl1998,VirbhadraEllis2000,BozzaEtAl2001,Bozza2002,HassePerlick2002,VirbhadraKeeton2008,Virbhadra2009,Bozza2010,Eiroa2012}. Near the critical orbit, the deflection angle takes a simple logarithmic form. From it, we can calculate the limiting angle, the separation of the outermost image, and the relative brightness \cite{BozzaScarpetta2007,BozzaMancini2004,Tsukamoto2018}. The divergent coefficient is controlled by the photon-sphere region, while the regular coefficient also depends on the outer part of the trajectory.

For the weak-deflection regime, the Gauss--Bonnet method provides a complementary geometric formulation in terms of the optical geometry \cite{GibbonsWerner2008,Werner2012}. For a spacetime that is not asymptotically flat, the source and the observer must be kept at finite distances \cite{RindlerIshak2007,Ishihara2016,OnoIshiharaAsada2017,OnoAsada2019}. Post-Minkowskian expansions and direct geodesic calculations are useful checks \cite{AdlerVirbhadra2022,HeEtAl2020}. In our model, the AdS and quintessence contributions preclude a naive limit at flat infinity. We therefore derive the exact finite-distance deflection angle and then use a locally matched weak-field slice for the analytical solar-limb comparison. The spacetime also incorporates perfect-fluid dark matter (PFDM), while the quintessence component follows Kiselev's static description of a black hole surrounded by a quintessence-like anisotropic source \cite{Kiselev2003}. PFDM has been investigated as an effective description motivated by galactic dynamics, in which the associated stress tensor introduces a logarithmic contribution to the lapse function \cite{RubinEtAl1980,MasseyEtAl2010,LiYang2012,ahmed2026shadow}. Its effects on black hole horizons, thermodynamics, particle motion, shadows, accretion, and gravitational lensing have been studied in both static and rotating geometries \cite{XuHouWang2018,MustafaHussain2021,FathiOlivaresVillanueva2022,MustafaEtAl2022,PantigOvgun2022,PantigEtAl2022,NampalliwarEtAl2021,XavierEtAl2023,PantigOvgun2023}. Recent studies have also explored configurations combining PFDM with quintessence or nonlinear electrodynamics (NLED), providing a broader context for investigating the interplay between environmental matter and modified electromagnetic fields \cite{HamilLutfuoglu2025,SuQiao2025,Romero-Figueroa:2026fan,ahmed2026shadow}.

The magnetic field contribution to the black hole solution considered in this work arises from NLED, a framework originally introduced to control the self-energy of point charges and later developed into an important tool for constructing exact black hole solutions \cite{Born1933,BornInfeld1934,HeisenbergEuler1936,Plebanski1970}. Their characteristic cones, causal properties, and effective photon metrics have been studied in detail \cite{Boillat1970,GibbonsRasheed1995,GibbonsHerdeiro2001,Novello2000,Novello2002,ObukhovRubilar2002,ShabadUsov2011,Schellstede2016,Sorokin2022}. Well-known magnetic and regular solutions were developed by Bronnikov, Ay\'on-Beato and Garc\'ia, Fan and Wang, Kruglov, and others \cite{Bronnikov2001,AyonBeatoGarcia1998,AyonBeatoGarcia1999,AyonBeatoGarcia2000,FanWang2016,Kruglov2017}. More generally, recent work has shown how energy conditions can help organize static, spherically symmetric effective geometries and families of regular spacetimes, including their horizons, photon spheres, and matching properties \cite{WangBattistaEnergy2026,WangBattistaFamilies2026}. Their shadows and lensing signals have also been studied in many recent papers, including several works by \"Ovg\"un and collaborators \cite{OkyayOvgun2022,Uniyal2023,UniyalEtAl2024,UniyalFathiOvgun2024,DePaulaEtAl2023,WaseemEtAl2025,PantigOvgun2025}.

The metric studied here combines a regularized magnetic mass term, a logarithmic PFDM term, quintessence, and an AdS scale. Its thermodynamics was introduced and developed in Refs.~\cite{Ndongmo2023,Gogoi2024}. Earlier studies also showed that shadows and bending angles require special care when the outer geometry is not flat \cite{Nam2018,BelmahiRbah2025}. We keep the present study focused. We follow the compact sequence used in the Lorentz-gauge lensing paper of \"Ovg\"un and Fathi \cite{OvgunFathi2025}, but every calculation is redone for the magnetic-PFDM-quintessence-AdS geometry.

For Hawking radiation, we use the Jacobi-metric tunnelling method. This method is closely related to the complex-path, Hamilton--Jacobi, and null-geodesic derivations \cite{SrinivasanPadmanabhan1999,ParikhWilczek2000,Angheben2005,Akhmedov2006,DiCriscienzoEtAl2007,VanzoAcquavivaDiCriscienzo2011,BeraGhoshMajhi2020,GibbonsJacobi2016,ChandaGibbonsGuha2017}. We show the main steps so that the origin of the imaginary part is clear and one can see why the simple zero of the lapse fixes the temperature.

To facilitate the comparison of the optical properties across different configurations, we introduce a one-parameter benchmark path connecting the Schwarzschild solution to a selected magnetic-PFDM-quintessence-AdS configuration. The parameter $\eta$ labels positions along this path and serves only as a convenient parameter for presenting the results; it does not represent a physical coupling. The analytical expressions are derived in terms of the original independent parameters, and $\eta$ is used exclusively in the figures, tables, and one-dimensional parameter intervals.

The paper is organized as follows. In Sec.~\ref{sec:model}, we introduce the magnetic black hole geometry, its parameters, the benchmark path, and the main curvature and horizon properties. In Sec.~\ref{sec:geodesics}, we analyze the photon geodesics, the effective potential, and the photon-sphere condition. In Sec.~\ref{sec:hawking}, we calculate the Hawking temperature using the Jacobi-metric tunnelling method and analyze the sparsity of the emitted radiation. Sec.~\ref{sec:weak} is devoted to weak gravitational lensing, where we use both the Gauss--Bonnet approach and a direct geodesic calculation. In Sec.~\ref{sec:shadow}, we study the photon sphere and the black hole shadow. Sec.~\ref{sec:strong} presents the strong-deflection limit and the corresponding lensing coefficients. In Sec.~\ref{sec:lens}, we discuss the relativistic images, Einstein rings, and simple observational intervals based on M87* and Sgr~A*. We summarize the main physical results in Sec.~\ref{sec:discussion} and give our final conclusions in Sec.~\ref{sec:conclusion}. The longer derivations, numerical checks, error propagation, and convergence tests are collected in the appendices. Throughout this work, we use $G=c=1$ and metric signature $(-,+,+,+)$.

\section{Magnetic AdS black hole in dark matter and quintessence}\label{sec:model}

\subsection{Spacetime geometry and model parameters}

The spacetime geometry is described by the following static, spherically symmetric line element\cite{Kiselev2003,LiYang2012,Ndongmo2023,Gogoi2024,HamilLutfuoglu2025}
\begin{equation}
\dd s^2=-B(r)\dd t'^2+\frac{\dd r^2}{B(r)}+r^2\left(\dd\theta^2+\sin^2\theta\,\dd\phi^2\right),
\label{eq:metric}
\end{equation}
with
\begin{equation}
B(r)=1-\frac{2Mr^2}{r^3+Q_m^3}
+\frac{\lambda}{r}\ln\left(\frac{r}{|\lambda|}\right)
-\frac{N}{r^{3w+1}}+\frac{r^2}{\ell^2}.
\label{eq:lapse}
\end{equation}
Here $M$ is the mass parameter, $Q_m$ controls the magnetic core, $\lambda$ characterizes the PFDM contribution, $N$ determines the strength of the quintessence field, and $\ell$ is the AdS radius. In the context of extended black-hole thermodynamics, the AdS scale can be promoted to a thermodynamic variable, providing a useful framework for studying the interplay between the geometry and thermodynamic properties of the spacetime; see, for instance, \cite{romero2024extended1,Ladino:2024ned}.
The black hole horizons are determined by the positive roots of $B(r_h)=0$. Since the logarithmic term makes this equation transcendental, the horizon radius must be obtained numerically \cite{ahmed2026shadow,Romero-Figueroa:2026fan}.
We focus on the outer horizon, which bounds the central static region of the black hole. Along the benchmark path defined in Eq.~\eqref{eq:path}, the horizon radius increases from $r_h=2M$ to $r_h\simeq2.1541M$. Since the sign of $B(r)$ determines the causal character of the temporal and radial coordinates, all optical calculations are restricted to the connected static region containing the observer. The asymptotic behavior of the lapse function also provides insight into the spacetime geometry. In the absence of PFDM ($\lambda=0$), a nonzero magnetic charge $Q_m$ replaces the Schwarzschild-like $1/r$ term with a contribution proportional to $-2Mr^2/Q_m^3$ near the origin, thereby removing the corresponding divergence in the lapse function. By contrast, the inclusion of the PFDM contribution reintroduces a curvature singularity at the origin through the logarithmic term \cite{ahmed2026shadow,Romero-Figueroa:2026fan}. 
Throughout this work, we adopt $w=-2/3$ for the quintessence field, so that its contribution to the lapse function takes the form $-Nr$. Moreover, at large radii, we obtain
\begin{equation}
 B(r)=1-Nr+\frac{r^2}{\ell^2}
 -\frac{2M}{r}+\frac{\lambda}{r}\ln\!\left(\frac{r}{|\lambda|}\right)
 +\frac{2MQ_m^3}{r^4}+O(r^{-7}).
 \label{eq:larger}
\end{equation}
\begin{figure}[ht!]
    \centering
    \begin{minipage}[t]{0.48\textwidth}
        \centering
        \includegraphics[width=\textwidth]{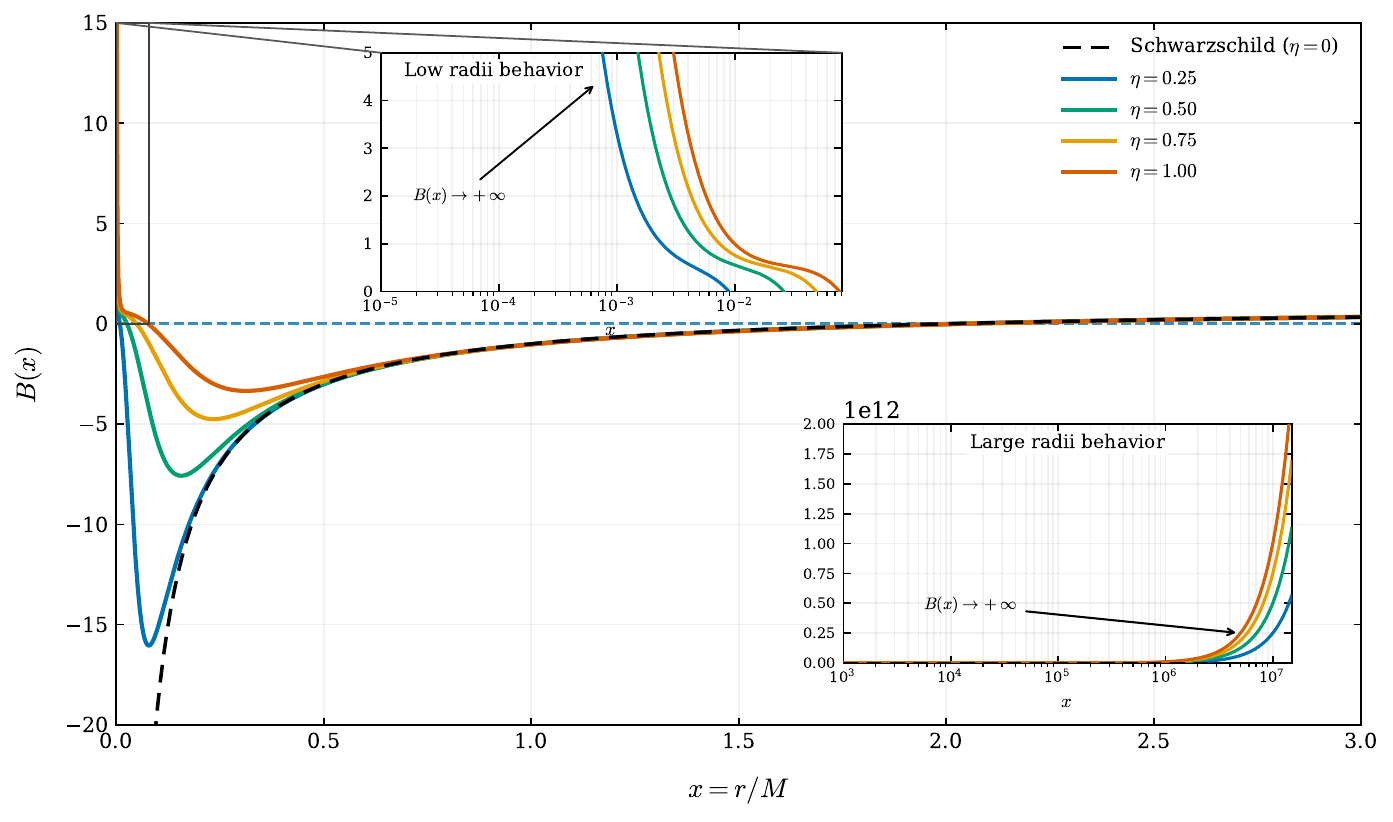}\\(a)
    \end{minipage}
    \hfill
    \begin{minipage}[t]{0.42\textwidth}
        \centering
        \includegraphics[width=\textwidth]{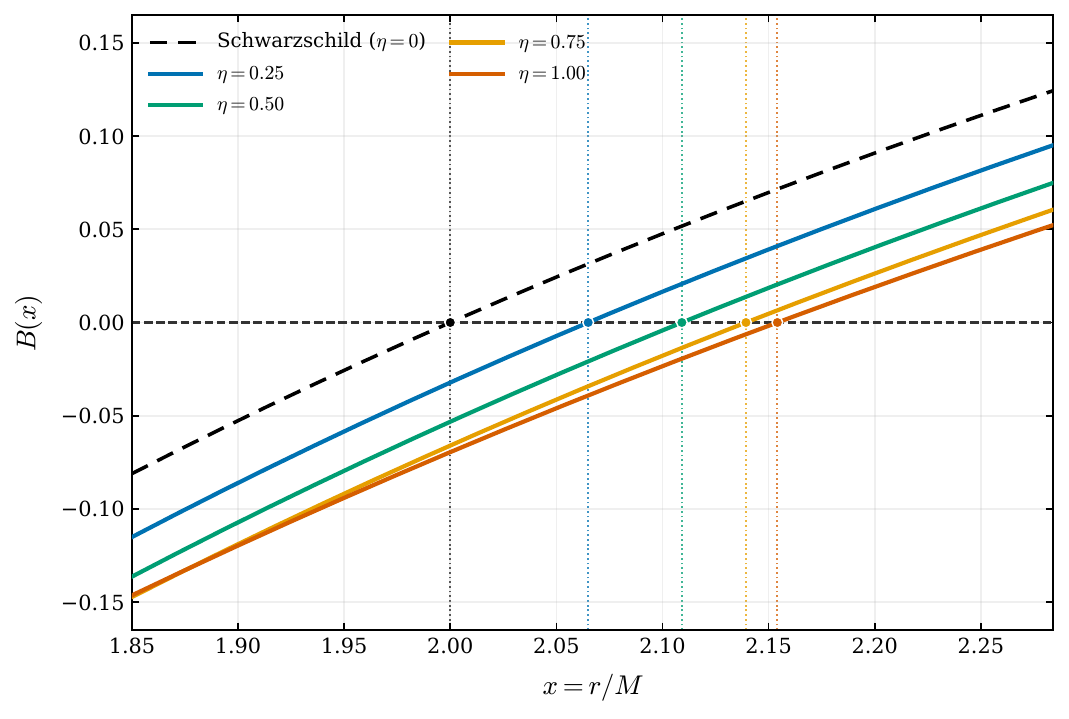}
        \\(b)
    \end{minipage}
    \caption{Dimensionless lapse function $B(x)$ for different values of $\eta$. Panel (a) shows the overall behavior, including the small- and large-radius insets. Panel (b) provides a zoom near the event horizons, with the horizon positions marked by points and vertical dotted lines. The horizontal dashed line indicates $B(x)=0$.}
    \label{fig:lapse}
\end{figure}

The leading terms $-Nr$ and $r^2/\ell^2$ govern the asymptotic behavior of the geometry, as illustrated in Fig.~\ref{fig:lapse}. This highlights why the lensing problem must be treated at finite distances or within a suitable matching prescription.
To further characterize the spacetime geometry, we consider the curvature invariants associated with the metric, Eq.~\eqref{eq:metric}. In particular, the Ricci scalar and the Kretschmann scalar follow from contractions of the Riemann tensor
\begin{equation}
 R=-B''-\frac{4B'}{r}+\frac{2(1-B)}{r^2},
 \label{eq:ricci}
\end{equation}
and
\begin{equation}
 K=R_{\mu\nu\rho\sigma}R^{\mu\nu\rho\sigma}
 =(B'')^2+4\left(\frac{B'}{r}\right)^2
 +4\left(\frac{1-B}{r^2}\right)^2.
 \label{eq:kretsch}
\end{equation}
Substitution of $B$ from Eq.~\eqref{eq:lapse} with $w=-2/3$ yields the compact Ricci scalar
\begin{equation}
 R=\frac{12MQ_m^3(2Q_m^3-r^3)}{(r^3+Q_m^3)^3}
 -\frac{\lambda}{r^3}+\frac{6N}{r}-\frac{12}{\ell^2}.
 \label{eq:ricciexplicit}
\end{equation}
This result separates the four contributions clearly. It also proves that the metric is not Schwarzschild written in different coordinates, because the scalar invariants change when any environmental parameter is nonzero. The term $\lambda/r^3$ shows the PFDM singularity at the center, while $-12/\ell^2$ is the constant AdS curvature. This distinction is useful here: a regularized magnetic core does not by itself make the full spacetime regular once the PFDM term is present \cite{ahmed2026shadow,Romero-Figueroa:2026fan}. Recent systematic studies of static spherical geometries also stress that regularity should be checked together with curvature invariants, energy conditions, and the global structure of the solution \cite{WangBattistaEnergy2026,WangBattistaFamilies2026}. To follow the invariant horizon test used in Ref.~\cite{OvgunFathi2025}, we also consider the induced two-dimensional metric
\begin{equation}
 \dd\tilde{\sigma}^2=B(r)\dd t^2+\frac{\dd r^2}{B(r)}.
 \label{eq:induced}
\end{equation}
Its curvature scalar is $-B''(r)$, so a convenient squared invariant is
\begin{equation}
 {}^{\Sigma}I_1=[B''(r)]^2.
 \label{eq:sI1}
\end{equation}
Taking gradients with the inverse induced metric gives
\begin{align}
 {}^{\Sigma}I_3&=B(r)[B'''(r)]^2,\label{eq:sI3}\\
 {}^{\Sigma}I_5&=4B(r)[B''(r)]^2[B'''(r)]^2.\label{eq:sI5}
\end{align}
Both gradient invariants vanish at a simple horizon because they contain the factor $B(r)$. They can also vanish at special extrema of the curvature. For this reason, we use them together with the roots and the sign of $B$, not as isolated horizon equations. The needed derivatives are listed in Appendix~\ref{app:derivatives}. The invariant test is coordinate independent, but the physical event horizon is still chosen from the causal structure of the spacetime. To formulate the horizon and curvature analysis in terms of a minimal set of parameters, we introduce the independent dimensionless combinations
\begin{equation}
 q=\frac{Q_m}{M},\qquad \lambda_M=\frac{\lambda}{M},\qquad n_q=NM,
 \qquad \ell_M=\frac{\ell}{M}.
 \label{eq:dimless}
\end{equation}

In terms of the dimensionless radial coordinate $x=r/M$, the lapse function in Eq.~\eqref{eq:lapse} becomes
\begin{equation}
 B(x)=1-\frac{2x^2}{x^3+q^3}
 +\frac{\lambda_M}{x}\ln\!\left(\frac{x}{|\lambda_M|}\right)
 -n_qx+\frac{x^2}{\ell_M ^2}.
 \label{eq:lapsedimless}
\end{equation}
This expression is convenient for the numerical analysis, as it is written entirely in terms of dimensionless quantities. To explore the transition from the Schwarzschild solution to the full configuration, we introduce the one-parameter path
\begin{equation}
 q=0.60\eta,\qquad
 \lambda_M=-0.050\eta,\qquad
 n_q=0.0020\eta,\qquad
 \ell_M^{-2}=\frac{\eta}{100^2},
 \qquad 0\leq\eta\leq1.
 \label{eq:path}
\end{equation}
At $\eta=0$, the lapse function reduces to the Schwarzschild form, whereas $\eta=1$ corresponds to the full environment considered. This parametrization defines a convenient one-dimensional path through parameter space without implying any physical relation among the parameters. 
As shown in Fig.~\ref{fig:lapse}(a), the lapse function exhibits two positive roots for the selected values of $\eta$, corresponding to an inner and an outer horizon. These roots delimit the region where $B(x)<0$. Fig.~\ref{fig:lapse}(b) provides a close-up of the outer horizon, showing its outward shift as $\eta$ increases along the chosen benchmark path.

\section{Photon geodesics and effective potential}
\label{sec:geodesics}

To study photon motion, we follow the standard Lagrangian formalism (see, e.g., Refs.~\cite{Chandrasekhar1983,cov04,FathiOlivaresVillanueva2022,Cruz:2013ufa}). We first consider a general static, spherically symmetric spacetime described by
\begin{equation}
    \dd s^2=-A(r)\dd t^2+\mathcal{B}(r)\dd r^2
    +C(r)\dd\Omega^2,
    \label{eq:ABC}
\end{equation}
where $\dd\Omega^2=\dd\theta^2+\sin^2\theta\,\dd\phi^2$.
The Lagrangian for geodesic motion is
\begin{equation}
    2\mathcal{L}
    =-A(r)\dot{t}^2+\mathcal{B}(r)\dot{r}^2
    +C(r)\dot{\theta}^2
    +C(r)\sin^2\theta\,\dot{\phi}^2
    =-\epsilon,
    \label{lagrangian r}
\end{equation}
where the dot denotes differentiation with respect to an affine parameter, $\epsilon=0$ for massless particles, and $\epsilon=1$ for massive particles. Since $t$ and $\phi$ are cyclic coordinates, their conjugate momenta are conserved:
\begin{equation}
    E=A(r)\dot{t},
    \qquad
    J=C(r)\sin^2\theta\,\dot{\phi}.
\end{equation}
By spherical symmetry, the motion can be restricted to an invariant plane, which we choose as $\theta=\pi/2$. For null geodesics with $E\neq0$, we define the impact parameter as
\begin{equation}
    b\equiv\frac{J}{E}.
    \label{eq:impactparameter}
\end{equation}
The null condition, $ds^2=0$, yields the radial equation
\begin{equation}
    A(r)\mathcal{B}(r)\dot{r}^2
    +V_{\rm eff}(r)=E^2,
    \qquad
    V_{\rm eff}(r)=J^2\frac{A(r)}{C(r)}.
    \label{eq:Veff}
\end{equation}
To characterize the optical properties of the spacetime, it is convenient to introduce the function $h(r)$ \cite{Synge1966,PerlickTsupkoBisnovatyi2015,PerlickTsupko2022},
\begin{equation}
    h(r)^2=\frac{C(r)}{A(r)}.
    \label{eq:hfun}
\end{equation}
This function relates the impact parameter to the radial turning points of null geodesics. In particular, at a radial turning point, where $\dot{r}=0$, we have $b^2=h(r)^2$. Furthermore, a circular photon orbit of radius $r_{\rm ph}$ corresponds to a constant radial coordinate and is therefore associated with an extremum of the function $h(r)^2$. For the metric functions specified in Eq.~\eqref{eq:metric}, namely, $A(r)\equiv B(r)$, $\mathcal{B}(r)\equiv 1/B(r)$, and $C(r)\equiv r^2$, the photon sphere condition becomes
\begin{equation}
\left.\left(h(r)^2\right)'\right|_{r=r_{\rm ph}}=0
    \qquad\Longrightarrow\qquad
    r_{\rm ph}B'(r_{\rm ph})-2B(r_{\rm ph})=0,
    \label{eq:photoncondition}
\end{equation}
where the prime denotes differentiation with respect to $r$. Moreover, the orbit is unstable if
\begin{equation}
    \left.
    \left(\frac{B(r)}{r^2}\right)''
    \right|_{r=r_{\rm ph}}<0.
    \label{eq:unstable}
\end{equation}
This stability check is important because some spacetimes can admit more than one circular photon orbit \cite{Cardoso2014}. The corresponding critical impact parameter is
\begin{equation}
    b_c=h(r_{\rm ph})
    =\frac{r_{\rm ph}}{\sqrt{B(r_{\rm ph})}}.
    \label{eq:bcgeneral}
\end{equation}
It separates photon trajectories captured by the black hole from those scattered back to infinity, thereby determining the critical boundary between capture and scattering and playing a key role in the black hole optical observables discussed in Secs.~\ref{sec:shadow}--\ref{sec:lens}.
We now apply this formalism to the spacetime described by the metric in Eq.~\eqref{eq:lapsedimless}.
Thus, in terms of the dimensionless radial coordinate $x=r/M$, our solution has
\begin{equation}
    A(x)=B(x),\qquad
    \mathcal{B}(x)=\frac{1}{B(x)},\qquad
    C(x)=x^2.
\end{equation}
The Lagrangian, Eq.~\eqref{lagrangian r}, becomes
\begin{equation}
    2\mathcal{L}
    =-B(x)\dot{t}^2+\frac{\dot{x}^2}{B(x)}
+x^2\dot{\theta}^2+x^2\sin^2\theta\,\dot{\phi}^2
    =-\epsilon,
\end{equation}
where $t=t'/M$ is the normalized time. Since $t$ and $\phi$ are cyclic coordinates, the corresponding conserved momenta are
\begin{eqnarray}
    \label{conserv11}
    \Pi_t
    &=&\frac{\partial\mathcal{L}}{\partial\dot{t}}
    =-B(x)\dot{t}\equiv-E,\\
    \label{conserv2}
    \Pi_{\phi}
    &=&\frac{\partial\mathcal{L}}{\partial\dot{\phi}}
    =x^2\sin^2\theta\,\dot{\phi}\equiv J.
\end{eqnarray}
Choosing the invariant plane $\theta=\pi/2$, the equations of motion read
\begin{eqnarray}
    \label{eqmoangx}
    \dot{\phi}&=&\frac{J}{x^2},\\
    \label{eqmorad2}
    \dot{x}^2&=&E^2-V(x).
\end{eqnarray}
For massless particles, the effective potential in Eq.~\eqref{eq:Veff} reduces to
\begin{equation}
    \label{effpotx}
    V(x)=\frac{J^2B(x)}{x^2}.
\end{equation}

\begin{figure}[ht!]
    \centering
    \begin{minipage}[t]{0.48\textwidth}
        \centering
        \includegraphics[width=\textwidth]{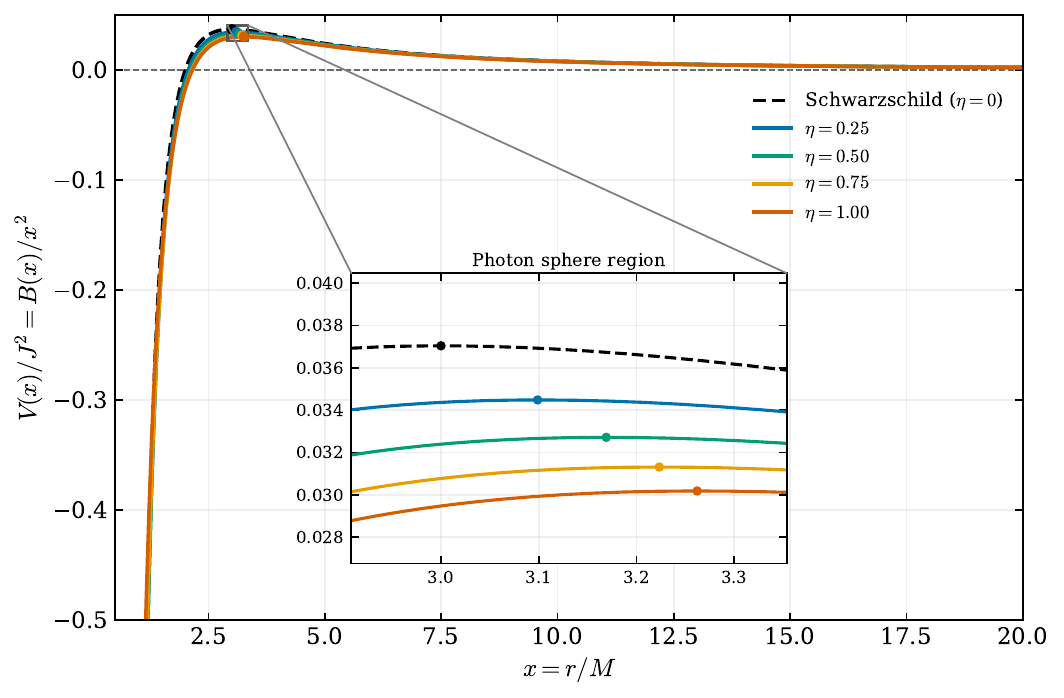}\\(a)
    \end{minipage}
    \hfill
    \begin{minipage}[t]{0.48\textwidth}
        \centering
        \includegraphics[width=\textwidth]{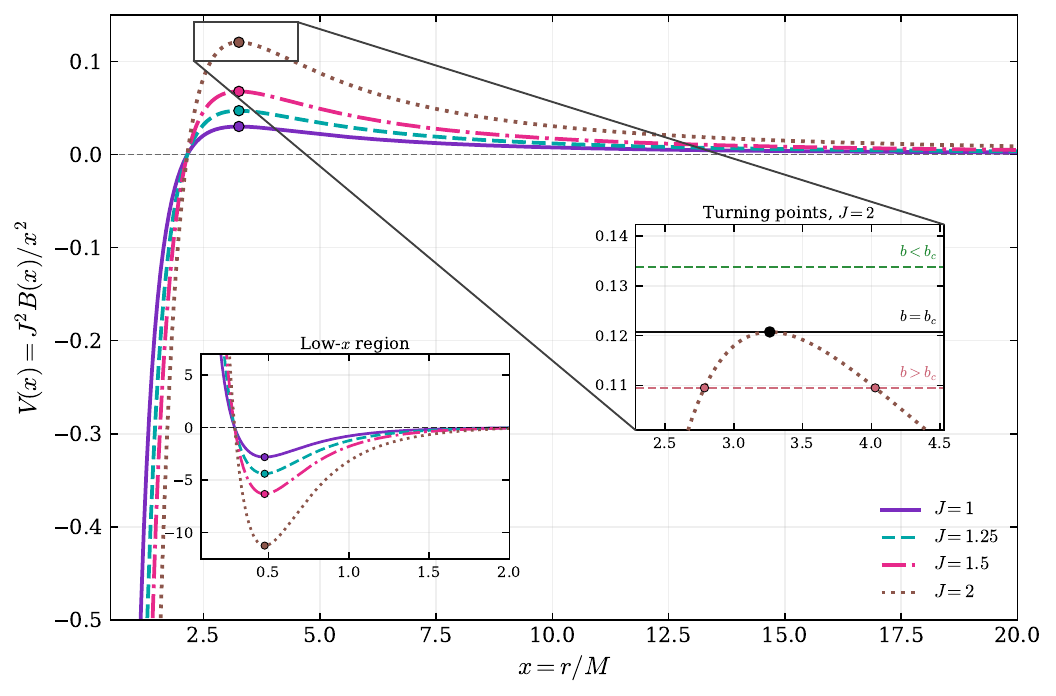}\\(b)
    \end{minipage}
    \caption{$V(x)$ for null geodesics: (a) normalized potential for different $\eta$, showing the photon-sphere region; (b) potential for fixed $\eta=1$ and different $J$, illustrating the turning points relative to $b_c$.}
    \label{fig:effect}
\end{figure}
The photon sphere condition in Eq.~\eqref{eq:photoncondition} takes the reduced form
\begin{equation}
    h(x)^2=\frac{x^2}{B(x)}
    \qquad\Longrightarrow\qquad
    \left.
    \left(
    x\frac{\dd B(x)}{\dd x}-2B(x)
    \right)
    \right|_{x=x_{\rm ph}}=0.
    \label{eqmin}
\end{equation}
\begin{table}[htbp]
    \centering
    \caption{Event-horizon radius $x_h$, photon-sphere radius $x_{\rm ph}$, and critical impact parameter $b_c$ for different values of $\eta$.}
    \label{tab:critical-photon}
    \begin{tabular}{cccc}
        \hline
        $\eta$ & $x_h$ & $x_{\rm ph}$ & $b_c$ \\
        \hline
        $0.00$ & $2.000000$ & $3.000000$ & $5.196152$ \\
        $0.25$ & $2.064985$ & $3.098841$ & $5.385235$ \\
        $0.50$ & $2.109119$ & $3.169024$ & $5.528650$ \\
        $0.75$ & $2.139335$ & $3.223310$ & $5.651174$ \\
        $1.00$ & $2.154133$ & $3.262004$ & $5.755947$ \\
        \hline
    \end{tabular}
\end{table}
In the benchmark range considered in Eq.~\eqref{eq:path}, the effective potential exhibits both an unstable and a stable circular photon orbit, as illustrated in Fig.~\ref{fig:effect}(b). The inner unstable orbit outside the event horizon determines the critical curve, whereas the stable orbit corresponds to the local minimum of the effective potential. The dimensionless critical impact parameter associated with the unstable photon orbit is given by
\begin{equation}
    b_c=\frac{x_{\rm ph}}{\sqrt{B(x_{\rm ph})}}.
    \label{eq:bcx}
\end{equation}
The event-horizon radii, photon-sphere positions, and critical impact parameters for the considered values of $\eta$ are reported in Table~\ref{tab:critical-photon}. Moreover, Fig.~\ref{fig:effect} shows the effective potentials for null geodesics. Fig~\ref{fig:effect}(a) displays the normalized potential $V(x)/J^2$ for different values of $\eta$. Although the curves are close to each other, the photon-sphere position shifts to the right as $\eta$ increases, relative to the Schwarzschild value $x_{\rm ph}=3$. Fig~\ref{fig:effect}(b) shows $V(x)$ for fixed $\eta=1$ and varying $J$. The photon-sphere position remains unchanged because varying $J$ only rescales $V(x)$.
We now turn to radial null geodesics to further examine photon propagation in this spacetime.


\subsection{Radial null geodesics $(J=0)$}
\label{subsecradialNull}

Radial null geodesics are particularly important for understanding the causal structure of the spacetime. For photons with $J=0$, the effective potential vanishes, $V(x)=0$. Accordingly, Eqs.~\eqref{conserv11} and \eqref{eqmorad2} yield
\begin{eqnarray}
    \dot{x}&=&\pm E,
    \label{mr.1}\\
    \frac{\dd x}{\dd t}&=&\pm B(x).
    \label{mr.2}
\end{eqnarray}
Choosing the initial radial position $x=x_i$ at $t=\tau=0$, these equations give
\begin{eqnarray}
    \tau(x)&=&\pm\frac{x-x_i}{E},
    \label{taunull}\\
    t(x)&=&\pm\int_{x_i}^{x}\frac{\dd x}{B(x)}.
    \label{tnull}
\end{eqnarray}

\begin{figure}[ht!]
    \centering
    \begin{minipage}[t]{0.5\textwidth}
        \centering
        \includegraphics[width=\textwidth]{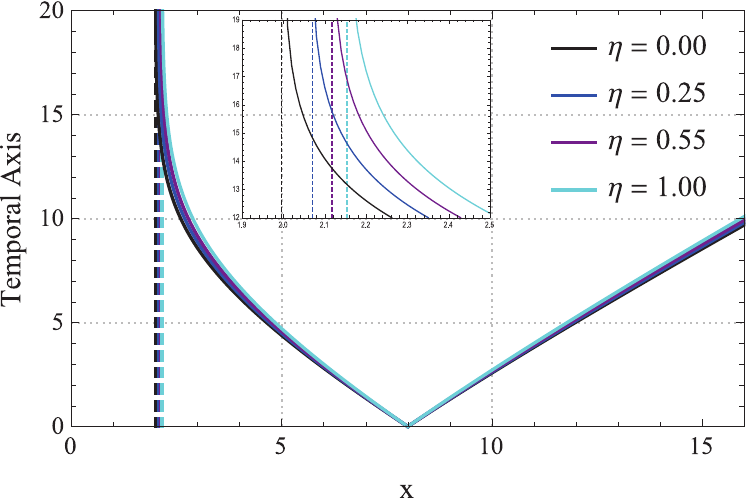}
        
    \end{minipage}
    \caption{Radial null geodesics, $\tilde{t}(x)$, for different values of the parameter $\eta$. The dashed, dotted, and solid curves correspond to $\eta=1$, $\eta=0.55$, and $\eta=0.25$, respectively.}\label{figtimenull} 
\end{figure}

Here, $\tau$ denotes the affine parameter for null geodesics rather than proper time. Eq.~\eqref{taunull} shows that photons cross a regular event horizon in a finite affine parameter and can subsequently reach the center at $r=0$, depending on the global structure of the spacetime \cite{Chandrasekhar1983,cov04,Cruz:2013ufa}. On the other hand, Eq.~\eqref{tnull} determines the coordinate time along the radial trajectories. Its evaluation requires integrating through the radial dependence of $B(x)$, 
which we solve numerically for different values of $\eta$, as shown in Fig.~\ref{figtimenull}. Consistent with previous studies of black-hole spacetimes \cite{Chandrasekhar1983,cov04,Cruz:2013ufa}, an external observer finds that photons approaching a regular horizon require an infinite coordinate time to reach it.

We next extend our analysis to nonradial photon motion, examining the possible types of motion in this gravitational background.

\subsection{Angular motion ($J\neq0$)}
\label{angmot}

\begin{figure}[ht!]
    \centering
    \includegraphics[width=60mm]{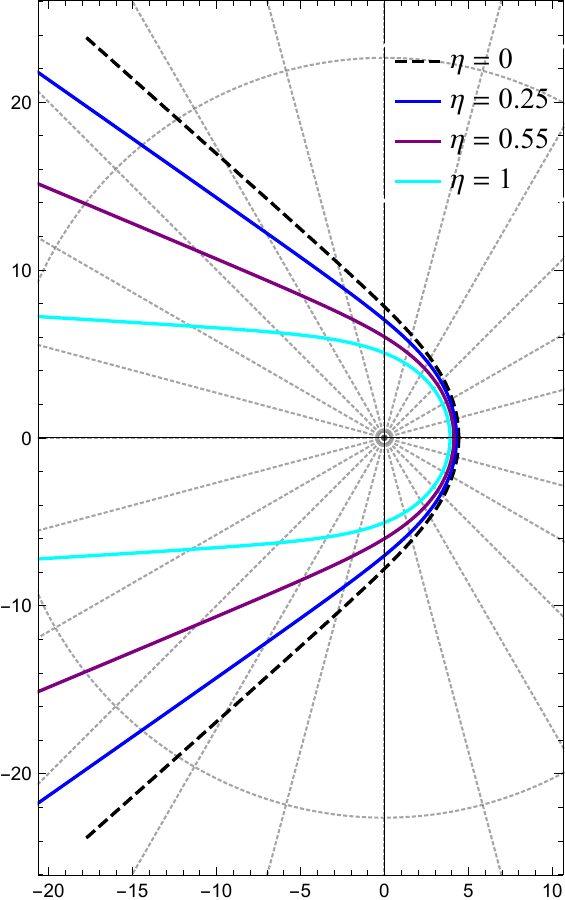}
    \caption{Polar plot for the deflection of light with $M=1$ and $J=10$. All orbits have the same energy, $E^2=2.8$.}
    \label{f3}
\end{figure}

To determine the angular trajectories of photons with $J\neq0$, we combine Eqs.~\eqref{eqmoangx}--\eqref{eqmorad2}. Introducing the impact parameter defined in Eq.~\eqref{eq:impactparameter}, we obtain the master equation
\begin{equation}
    \label{eqmotL2}
    \left(\frac{\dd x}{\dd\phi}\right)^2
    =\frac{x^4}{b^2}-x^2B(x).
\end{equation}

\begin{figure}[ht!]
    \centering
    \includegraphics[width=50mm]{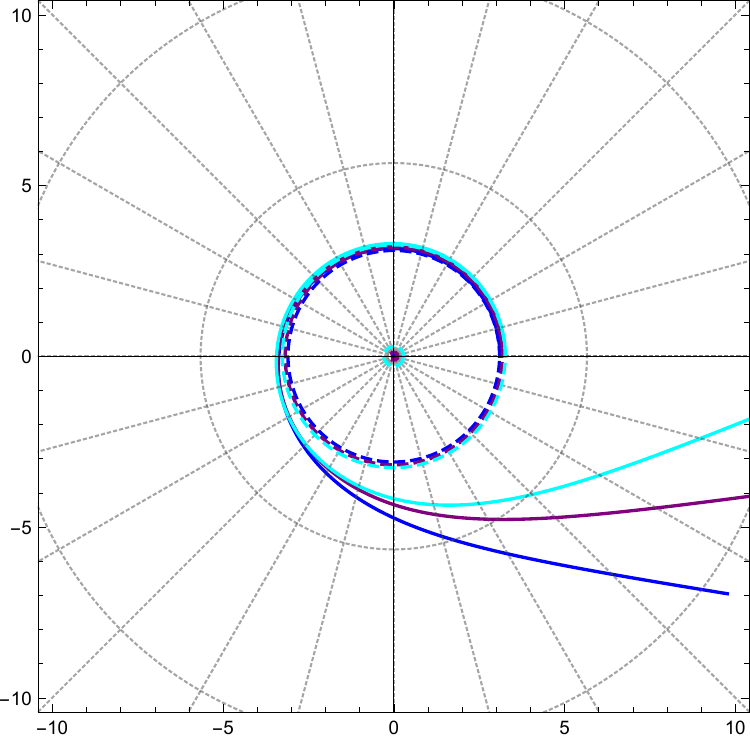}\qquad
    \includegraphics[width=50mm]{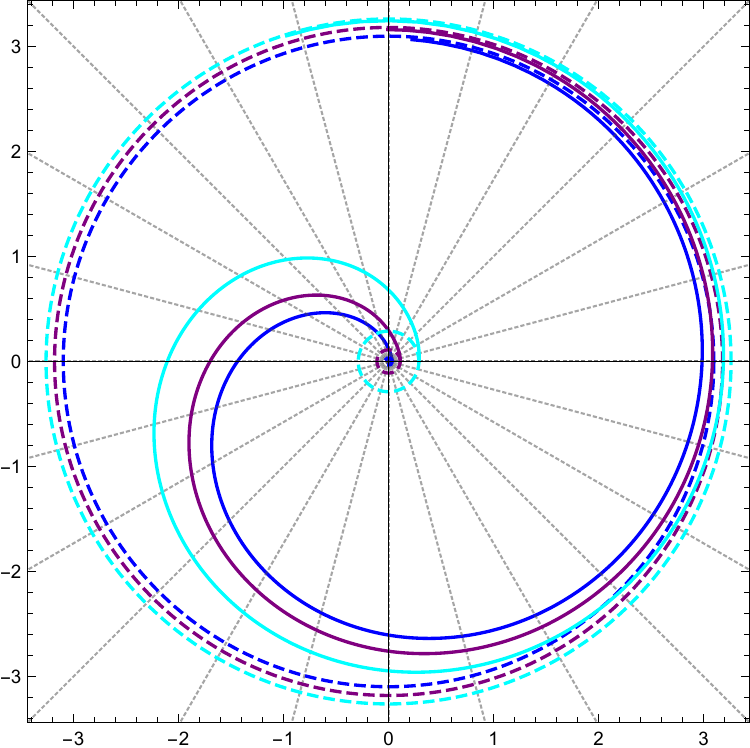}
    \caption{Critical orbits for $M=1$ and $J=10$. The left panel shows COFK, while the right panel shows COSK.}
    \label{criticas}
\end{figure}

\begin{figure}[ht!]
    \centering
    \begin{minipage}{0.48\textwidth}
        \centering
        \includegraphics[width=60mm]{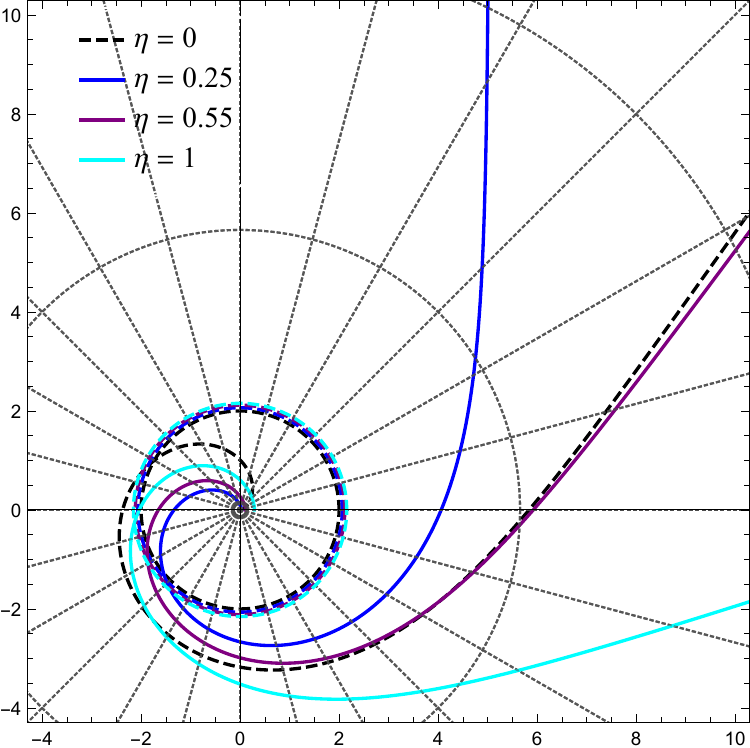}
        \par\smallskip
        (a)
    \end{minipage}
    \hfill
    \begin{minipage}{0.48\textwidth}
        \centering
        \includegraphics[width=60mm]{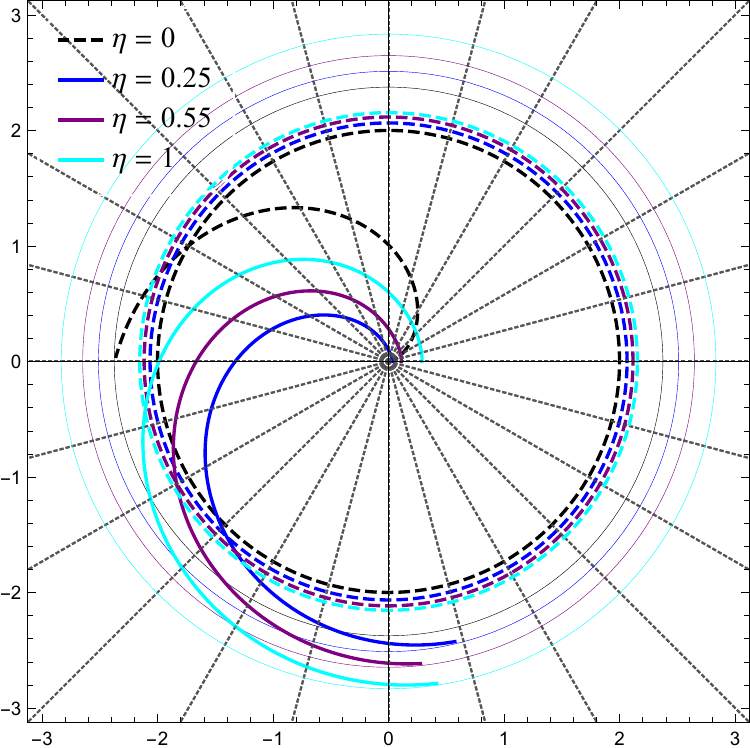}
        \par\smallskip
        (b)
    \end{minipage}
    \caption{(a) Capture zone for photons with $M=1$, $J=10$, and $E^2=4$. (b) Polar plot for orbits of the second kind with $M=1$, $J=10$, and $E^2=2.8$.}
    \label{fig:photon-trajectories-captura}
\end{figure}

This equation determines the possible photon trajectories and their radial turning points. The angular trajectory follows by integrating Eq.~\eqref{eqmotL2},
\begin{equation}
    \phi(x)=\pm b\int_{x_i}^{x}
    \frac{\dd x}{\sqrt{x^4-b^2x^2B(x)}}.
    \label{eqmotL3}
\end{equation}
Based on the impact parameter and the effective potential in Eq.~\eqref{effpotx}, we identify the different types of photon trajectories, illustrated in Figs.~\ref{f3}, \ref{criticas}, and \ref{fig:photon-trajectories-captura}.

\paragraph{Deflection zone.}

Orbits of the first kind are allowed for $b>b_c$, provided that an accessible turning point $x_d>x_{ph}$ exists. Here, $x_d$ is determined by $V(x_d)=E^2$. Photons arriving from infinity are deflected by the gravitational field, reach the turning point, and return to infinity. As the impact parameter approaches the critical value from above, the turning point approaches the photon sphere, and the deflection angle increases. An example of these trajectories is shown in Fig.~\ref{f3}, where the photon paths are plotted in polar coordinates. The curves illustrate how the gravitational field bends the trajectories, with photons passing closer to the photon sphere experiencing stronger deflection.

\paragraph{Critical orbits.}

For $b=b_c$, the turning point coincides with the unstable circular photon orbit at $x=x_{ph}$. Photons approaching this orbit asymptotically complete an increasing number of revolutions around the black hole. Two classes of critical trajectories can be distinguished: critical orbits of the first kind (COFK), in which photons approach the photon sphere from outside, and critical orbits of the second kind (COSK), in which photons approach it from radii smaller than $x_{ph}$. Both classes are illustrated in Fig.~\ref{criticas}. The figure shows the asymptotic behavior of the trajectories near the unstable circular orbit, which separates scattering trajectories from those associated with capture by the black hole. The affine-parameter period and the coordinate-time period for a complete revolution along the circular orbit are
\begin{equation}
    T_{\lambda}=\frac{2\pi x_{ph}^2}{J},
    \qquad
    T_t=2\pi b_c
    =\frac{2\pi x_{ph}}{\sqrt{B(x_{ph})}}.
\end{equation}

\paragraph{Capture zone.}

Photons incident from infinity with an impact parameter smaller than the critical value, $b<b_c$, have no accessible turning point outside the photon sphere and are captured by the black hole. Their trajectories cross the photon-sphere region and continue toward the event horizon. The corresponding dimensionless capture cross section is
\begin{equation}
    \sigma_{cap}=\pi b_c^2
    =\frac{\pi x_{ph}^2}{B(x_{ph})}.
\end{equation}
An example of photon trajectories in the capture zone is shown in Fig.~\ref{fig:photon-trajectories-captura}(a). The plotted paths illustrate how photons with subcritical impact parameters are unable to escape back to infinity and instead plunge into the black hole.

\paragraph{Orbits of the second kind.}

Photons initially located between the event horizon and the photon sphere, $x_h<x<x_{ph}$, can also follow trajectories directed toward the black hole. In this region, photons moving inward cross the event horizon rather than returning to the exterior. An example is shown in Fig.~\ref{fig:photon-trajectories-captura}(b), where the trajectory illustrates the inward motion from the region inside the photon sphere toward the horizon. These trajectories complement the critical and capture cases by describing photon motion on the inner side of the photon sphere.

\section{Hawking radiation from the Jacobi metric}\label{sec:hawking}

A stationary black hole has entropy set by its horizon area and temperature set by its surface gravity \cite{Bekenstein1973,Hawking1975}. The tunnelling picture gives a local semiclassical derivation of this temperature. An outgoing mode crosses the horizon and acquires an imaginary part in its action \cite{SrinivasanPadmanabhan1999,ParikhWilczek2000,Angheben2005,Akhmedov2006,VanzoAcquavivaDiCriscienzo2011}. We use the Jacobi metric because it places the conserved particle energy directly in the spatial geometry \cite{GibbonsJacobi2016,ChandaGibbonsGuha2017,BeraGhoshMajhi2020}. For a particle of mass $m$, the Hamilton--Jacobi equation is
\begin{equation}
 g^{\mu\nu}\partial_\mu S\,\partial_\nu S+m^2=0.
 \label{eq:HJmass}
\end{equation}
Because the metric is stationary, we write $S=-Et+S_J(x^i)$. Solving the Hamilton--Jacobi equation for the spatial part gives
\begin{equation}
 \dd s_J^2=\left[E^2-m^2B(r)\right]
 \left[\frac{\dd r^2}{B(r)^2}+\frac{r^2}{B(r)}
 \left(\dd\theta^2+\sin^2\theta\,\dd\phi^2\right)\right].
 \label{eq:jacobi}
\end{equation}
This is the Jacobi line element. The same expression follows from the reduced action, and it is positive in the classically allowed spatial region. For radial motion,
\begin{equation}
 p_r=\partial_rS=\mp\frac{\sqrt{E^2-m^2B(r)}}{B(r)}.
 \label{eq:pr}
\end{equation}
The two signs describe the two radial directions. The only part that can produce an imaginary contribution is the simple pole at a nonextremal horizon. Let $r_h$ be a simple root of $B$. Close to the horizon,
\begin{equation}
 B(r)=B'(r_h)(r-r_h)+O[(r-r_h)^2]
 =2\kappa(r-r_h)+O[(r-r_h)^2],
 \label{eq:nearhorizon}
\end{equation}
where
\begin{equation}
 \kappa=\frac12B'(r_h)
 \label{eq:kappa}
\end{equation}
\begin{figure}[ht!]
 \centering
 \includegraphics[width=0.55\textwidth]{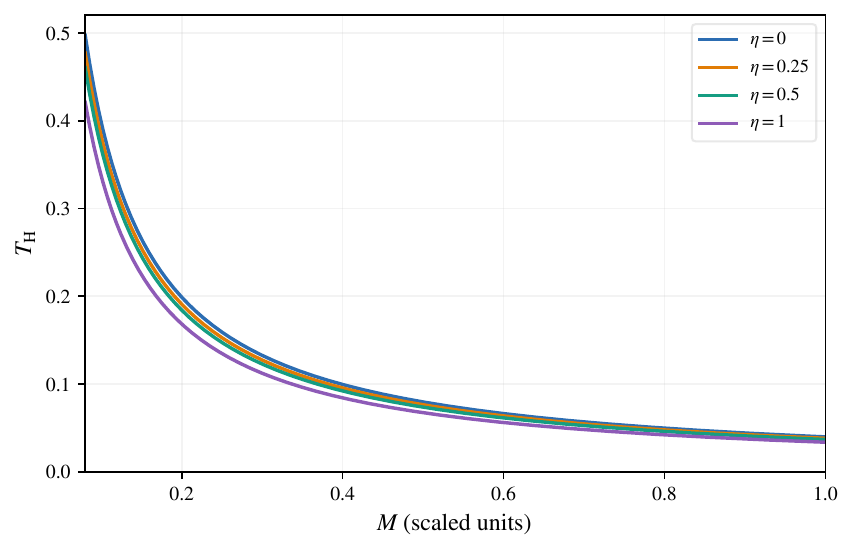}
 \caption{Hawking temperature as a function of the mass scale for representative values of the benchmark parameter $\eta$. The point $\eta=0$ is Schwarzschild.}
 \label{fig:temperature}
\end{figure}
is the surface gravity. Expanding the square root in the radial momentum gives
\begin{equation}
 p_r=\mp\left[\frac{E}{2\kappa(r-r_h)}-\frac{m^2}{2E}+O(r-r_h)\right].
 \label{eq:prexpansion}
\end{equation}
The mass term is regular, so it does not change the tunnelling exponent. A small contour around the pole gives
\begin{equation}
 \operatorname{Im}S_{\rm out}=\frac{\pi E}{2\kappa},\qquad
 \operatorname{Im}S_{\rm in}=-\frac{\pi E}{2\kappa}.
 \label{eq:imS}
\end{equation}
We then take the ratio of the outgoing and incoming probabilities. This avoids the usual factor-of-two ambiguity of a one-sided amplitude \cite{Angheben2005,Akhmedov2006,VanzoAcquavivaDiCriscienzo2011}:
\begin{equation}
 \Gamma=\frac{P_{\rm out}}{P_{\rm in}}
 =\exp\!\left[-\frac{2}{\hbar}
 \left(\operatorname{Im}S_{\rm out}-\operatorname{Im}S_{\rm in}\right)\right]
 =\exp\!\left(-\frac{2\pi E}{\hbar\kappa}\right).
 \label{eq:gamma}
\end{equation}
Comparing this expression with $\exp(-E/T_H)$ gives
\begin{equation}
 T_H=\frac{\hbar\kappa}{2\pi}=\frac{\hbar B'(r_h)}{4\pi}.
 \label{eq:THgeneral}
\end{equation}

We set $\hbar=1$ from now on. The same result follows from covariant Hamilton--Jacobi and null-geodesic methods \cite{ParikhWilczek2000,DiCriscienzoEtAl2007,VanzoAcquavivaDiCriscienzo2011}. For $w=-2/3$, the derivative listed in Appendix~\ref{app:derivatives} gives
\begin{align}
 T_H=\frac{1}{4\pi}\Bigg[
 &\frac{2Mr_h(r_h^3-2Q_m^3)}{(r_h^3+Q_m^3)^2}
 +\frac{\lambda}{r_h^2}\left(1-\ln\frac{r_h}{|\lambda|}\right)
 -N+\frac{2r_h}{\ell^2}\Bigg].
 \label{eq:THfull}
\end{align}
We may also use the horizon equation to remove $M$,
\begin{equation}
 M=\frac{r_h^3+Q_m^3}{2r_h^2}
 \left[1+\frac{\lambda}{r_h}\ln\!\left(\frac{r_h}{|\lambda|}\right)
 -Nr_h+\frac{r_h^2}{\ell^2}\right].
 \label{eq:Mhorizon}
\end{equation}

The first form keeps the mass explicit. The second expresses the temperature through the horizon and the external parameters. Both reduce to $T_H=1/(8\pi M)$ in the Schwarzschild limit. Fig.~\ref{fig:temperature} shows the Hawking temperature along the benchmark path. We keep the dimensionless geometry fixed and change only the overall mass scale, so all curves follow the expected $M^{-1}$ behavior. For the selected negative-PFDM benchmark, a larger $\eta$ lowers the dimensionless surface gravity.

Appendix~\ref{app:jacobi} gives the remaining steps between the mass-shell equation, the Jacobi metric, and the contour around the horizon. With the Hawking temperature determined from the near-horizon tunnelling analysis, we can now characterize not only the energy scale of the emitted quanta but also the temporal dilution of the radiation. This motivates the following analysis of the Hawking sparsity.

\subsection{Sparsity of Hawking radiation}\label{sec:sparsity}
\begin{figure}[ht!]
    \centering
    \includegraphics[width=0.8\textwidth]{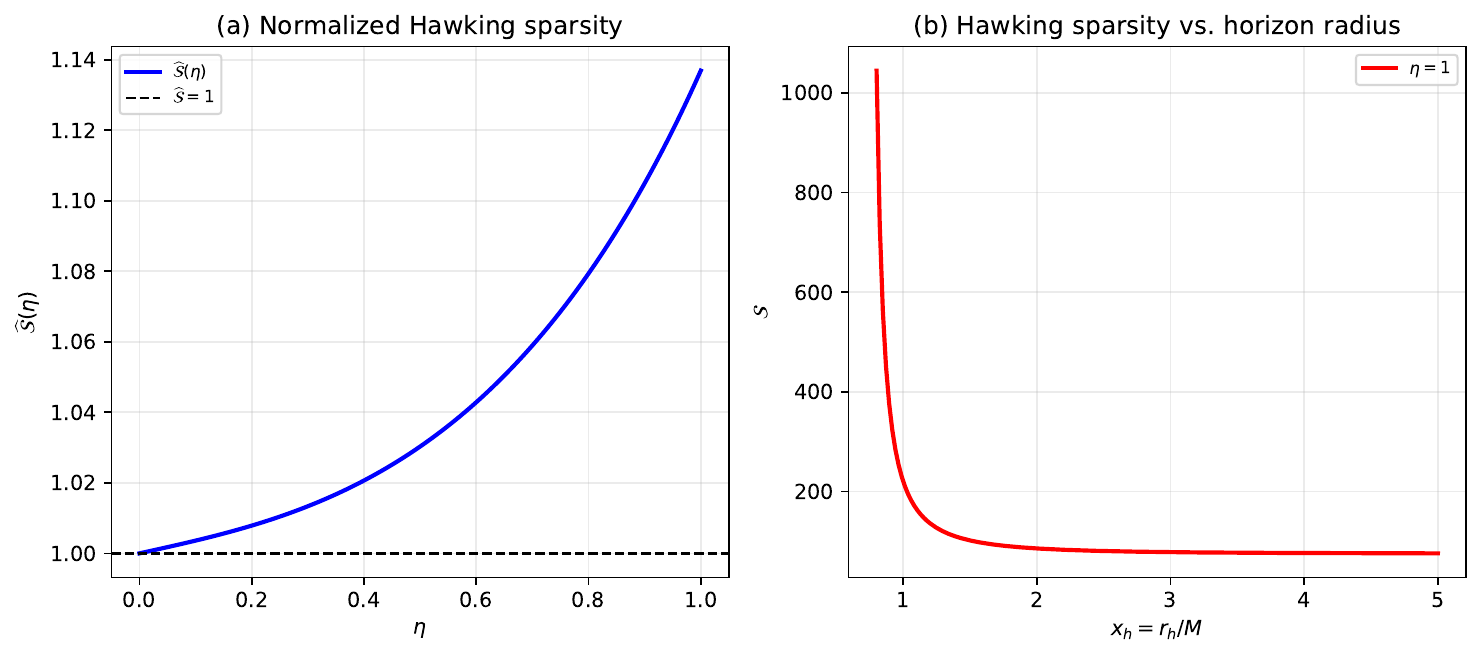}
    \caption{Normalized Hawking sparsity $\widehat{\mathcal{S}}=\mathcal{S}/\mathcal{S}_{\rm Sch}$
    as a function of the benchmark parameter $\eta$ (left panel), and
    Hawking sparsity $\mathcal{S}$ as a function of the dimensionless
    horizon radius $x_h=r_h/M$ for the full configuration $\eta=1$
    (right panel). The dashed line in the left panel indicates the
    Schwarzschild value, $\widehat{\mathcal{S}}=1$.}
    \label{fig:sparsity_eta_xh}
\end{figure}
The Hawking temperature sets the characteristic energy scale of the emitted radiation, while the emission rate, determined by the thermal spectrum and the effective emitting area, controls the characteristic temporal separation between successive Hawking quanta \cite{gray2016hawking}. In this section, we investigate the sparsity of the Hawking radiation emitted by the magnetic AdS black hole in the presence of PFDM and quintessence. The sparsity characterizes the temporal dilution of the emission process by comparing the characteristic wavelength of the thermal radiation with an effective emitting area \cite{gray2016hawking}. For a black hole with Hawking temperature $T_H$, the characteristic thermal wavelength is $\lambda_T=2\pi/T_H$. The sparsity parameter is defined as \cite{al2025shadow,baruah2025quasinormal}
\begin{equation}
    \mathcal{S}
    =
    \frac{\mathcal{C}}{\tilde{g}}
    \frac{\lambda_T^2}{A_{\rm eff}},
    \label{eq:sparsity_definition}
\end{equation}

where $\mathcal{C}$ is a dimensionless numerical factor,
$\tilde{g}$ denotes the spin degeneracy of the emitted quanta, and
$A_{\rm eff}$ is the effective emitting area \cite{al2025shadow,gray2016hawking}. For the spherically symmetric geometry in Eq.~\eqref{eq:metric},
following \cite{gray2016hawking}, we estimate the effective emitting
area from the characteristic photon-capture scale set by the critical
impact parameter as
$  A_{\rm eff}=4\pi b_c^2$, where $b_c$ is the critical impact parameter defined in
Sec.~\ref{sec:geodesics}, Eq.~\eqref{eq:impactparameter}. Hence, the sparsity parameter read
\begin{equation}
    \mathcal{S}
    =
    \frac{\mathcal{C}}{\tilde{g}}
    \frac{\pi}
    {b_c^2T_H^2}.
    \label{eq:S_TH}
\end{equation}
Moreover, substituting the Hawking temperature given by
Eq.~\eqref{eq:THgeneral} into Eq.~\eqref{eq:S_TH}, and using the
dimensionless variables and benchmark path introduced in
Eqs.~\eqref{eq:dimless} and \eqref{eq:path}, respectively, allows the
sparsity parameter along the benchmark path to be expressed directly
in terms of the radial derivative of the dimensionless metric function
at the outer horizon as
\begin{equation}
    \mathcal{S}(\eta)
    =
    \frac{\mathcal{C}}{\tilde{g}}
    \frac{16\pi^3}
    {{b}_c(\eta)^2
    \left[
    \partial_x B(x_h;\eta)
    \right]^2}.
    \label{eq:S_benchmark}
\end{equation}
Here, $x_h(\eta)$ is the outer horizon obtained from
$B(x_h;\eta)=0$, while $ b_c(\eta)$ is determined from
Eq.~\eqref{eq:bcx}. Thus, along the benchmark path defined in Eq.~\eqref{eq:path}, the sparsity
parameter becomes explicitly
\begin{equation}
\mathcal{S}(\eta)
=
\frac{\mathcal{C}}{\tilde{g}}
\frac{16\pi^3}
{b_c(\eta)^2
\left[
\displaystyle
\frac{2x_h(\eta)
\left\{
x_h(\eta)^3-2[0.60\eta]^3
\right\}}
{\left\{
x_h(\eta)^3+[0.60\eta]^3
\right\}^2}
+
\frac{0.050\eta}{x_h(\eta)^2}
\left[
\ln\left(
\frac{x_h(\eta)}{0.050\eta}
\right)-1
\right]
-0.0020\eta
+\frac{2\eta x_h(\eta)}{100^2}
\right]^2
}.
\label{eq:S_benchmark_explicit}
\end{equation}
\begin{figure}[ht!]
    \centering
    \includegraphics[width=0.5\textwidth]{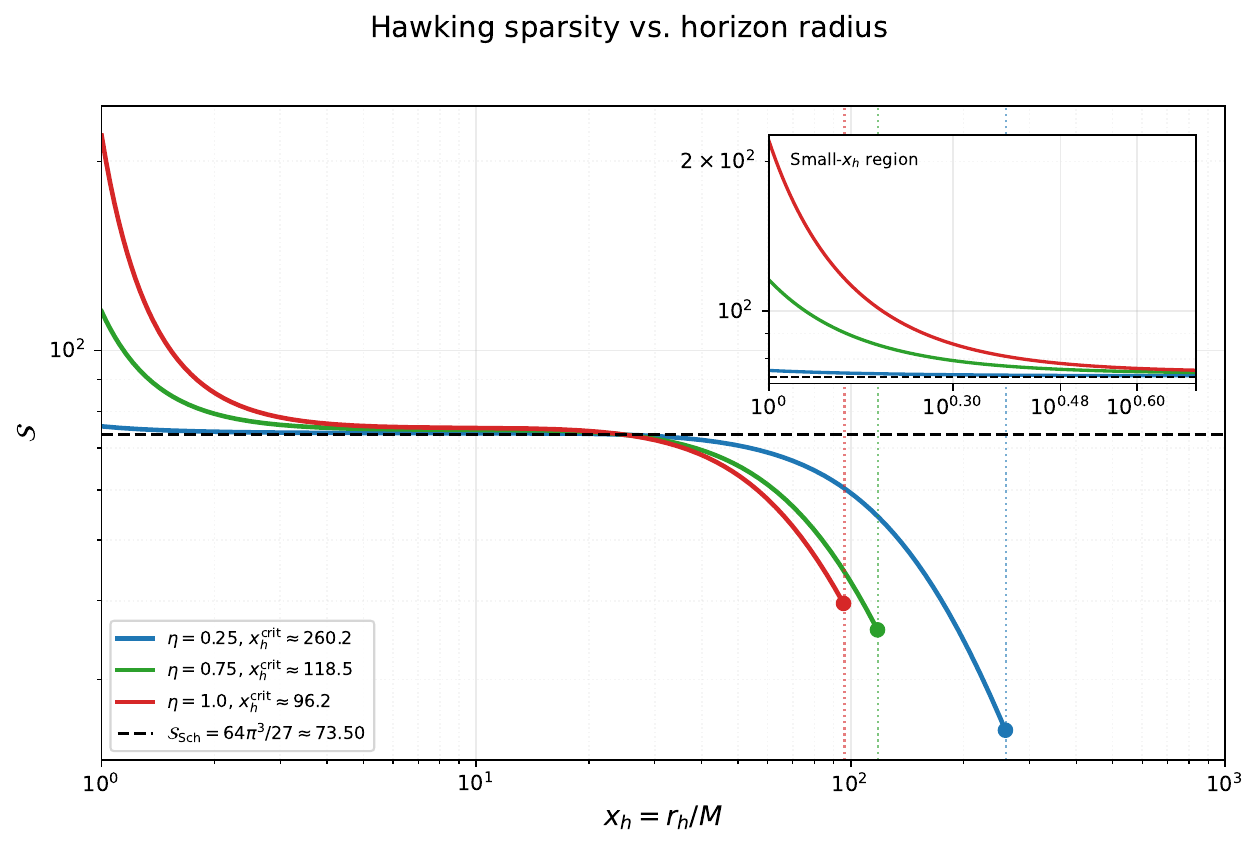}
    \caption{Hawking sparsity $\mathcal{S}$ as a function of the $x_h=r_h/M$ for $\eta=0.25$, $0.75$, and $1$. The horizontal dashed line denotes the Schwarzschild reference value $\mathcal{S}_{\rm Sch}\simeq73.50$, while the vertical dotted lines indicate the critical horizon radii $x_h^{\rm crit}$ beyond which no unstable photon sphere is found. The inset shows a magnified view of the small-$x_h$ region.}
    \label{fig:sparsity_eta_xh2}
\end{figure}
\begin{figure}[ht!]
    \centering
    \includegraphics[width=1\textwidth]{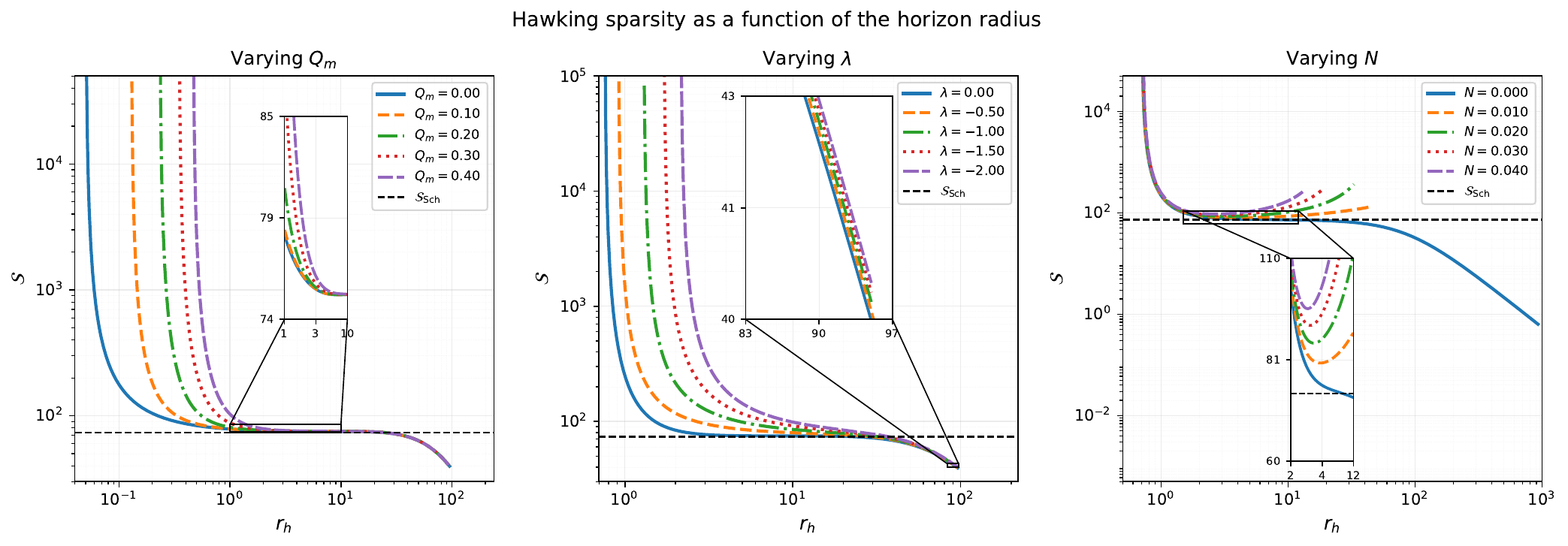}
    \caption{Hawking sparsity $\mathcal{S}$ as a function of $r_h$ for different values of the magnetic charge $Q_m$, PFDM parameter $\lambda$, and quintessence parameter $N$, while keeping the remaining parameters fixed at their benchmark values.}
    \label{fig:sparsity_parameter_dependence}
\end{figure}
For comparison, adopting the conventional normalization
$\mathcal{C}=\tilde{g}=1$, corresponding to a single spin state of the
emitted massless bosonic quanta \cite{gray2016hawking}, the Schwarzschild
limit, $\eta=0$, gives
\begin{equation}
     b_c=3\sqrt{3},
    \qquad
    x_h=2,
    \qquad
    \left.\partial_xB\right|_{x_h}=\frac{1}{2}
    \quad\Longrightarrow\quad
    \mathcal{S}_{\rm Sch}
    =
    \frac{64\pi^3}{27}\simeq 73.5\gg1.
    \label{eq:S_Sch}
\end{equation}
This large value reflects the highly sparse character of Hawking radiation, in contrast to ordinary black-body emitters, for which $\mathcal{S}\ll1$ \cite{alonso2018generalized}.
\begin{table}[t]
\centering
\caption{Dimensionless Hawking temperature and sparsity parameter along the benchmark path defined by Eq.~\eqref{eq:path}.}
\label{tab:sparsity_benchmark}
\begin{tabular}{c c c}
\hline
$\eta$ & $M T_H$ & $\mathcal{S}$ \\
\hline
0.00 & 0.039789 & 73.4964 \\
0.25 & 0.038193 & 74.2624 \\
0.50 & 0.036842 & 75.2309 \\
0.75 & 0.035392 & 78.5360 \\
1.00 & 0.033687 & 83.5587 \\
\hline
\end{tabular}
\end{table}
In addition, to quantify deviations from the Schwarzschild black hole case, we define the normalized sparsity
\begin{equation}
    \widehat{\mathcal{S}}(\eta)
    \equiv
    \frac{\mathcal{S}(\eta)}
    {\mathcal{S}_{\rm Sch}}
    =
    \frac{27}
    {4 b_c(\eta)^2
    \left[\partial_xB(x_h;\eta)\right]^2}.
    \label{eq:S_normalized}
\end{equation}
\begin{table}[tbp]
    \centering
    \caption{Critical horizon and photon-sphere radii for the considered values of $\eta$. A dash indicates that no unstable photon-sphere root was identified at the critical point.}
    \label{tab:critical_radii}
    \begin{tabular}{ccc}
        \hline
        $\eta$ & $x_h^{\rm crit}$ & $x_{\rm ph}^{\rm crit}$ \\
        \hline
        $0.25$ & $260.180$ & $1999.461$ \\
        $0.75$ & $118.482$ & $--$ \\
        $1.00$ & $96.165$  & $499.786$ \\
        \hline
    \end{tabular}
\end{table}
Consequently, $\widehat{\mathcal{S}}(0)=1$, while deviations from unity quantify modifications of the Hawking sparsity relative to the Schwarzschild case. As shown in Fig.~\ref{fig:sparsity_eta_xh}, $\mathcal{S}$ increases from the Schwarzschild value $\mathcal{S}_{\rm Sch}\simeq73.49$ at $\eta=0$ to $\mathcal{S}\simeq83.56$ at $\eta=1$, corresponding to a $13.7\%$ increase. This behavior is directly associated with the decrease in temperature as $\eta$ increases, as depicted in Fig.~\ref{fig:temperature} and summarized in Table~\ref{tab:sparsity_benchmark}. In addition,
Fig.~\ref{fig:sparsity_eta_xh2} illustrates the gradual approach of the
Hawking sparsity toward the Schwarzschild reference value
$\mathcal{S}_{\rm Sch}$ at intermediate horizon radii. For sufficiently
large $x_h$, the sparsity departs from this reference behavior, and the
corresponding branches terminate at a critical horizon radius
$x_h^{\rm crit}$ due to the loss of the unstable photon sphere,
determined by Eqs.~\eqref{eq:photoncondition}--\eqref{eq:unstable}.

The corresponding critical horizon and photon-sphere radii are
summarized in Table~\ref{tab:critical_radii}. Finally, Fig.~\ref{fig:sparsity_parameter_dependence} shows that the different sectors systematically enhance the Hawking sparsity as $Q_m$, $|\lambda|$, or $N$ increases, with the $N$-sector exhibiting the most pronounced effect. This stronger sensitivity to $N$, especially at intermediate and large horizon radii, highlights its significant impact on the Hawking emission \cite{al2025shadow}. For the $Q_m$- and $|\lambda|$-sectors, the Hawking sparsity decreases monotonically with increasing horizon radius for all parameter choices. In contrast, increasing $N$ progressively modifies this behavior: the sparsity can increase for intermediate-size black holes, producing a non-monotonic dependence on $r_h$, before eventually decreasing at larger horizon radii.

\section{Weak deflection angle}\label{sec:weak}

We investigate weak gravitational lensing in the magnetic AdS black hole spacetime, focusing on the effects of PFDM and quintessence on light propagation.

\subsection{Optical geometry and finite-distance definition}

On the equatorial plane, a null ray satisfies
\begin{equation}
 \dd t^2=\frac{\dd r^2}{B(r)^2}+\frac{r^2}{B(r)}\dd\phi^2.
 \label{eq:optical}
\end{equation}
For a two-dimensional metric $\dd t^2=E(r)\dd r^2+G(r)\dd\phi^2$, the Gaussian curvature is
\begin{equation}
 \cK=-\frac{1}{\sqrt{EG}}\frac{\dd}{\dd r}
 \left[\frac{1}{\sqrt E}\frac{\dd\sqrt G}{\dd r}\right].
 \label{eq:Kgeneral}
\end{equation}
Here $E=B^{-2}$ and $G=r^2/B$, so
\begin{equation}
 \cK=\frac12\left[B(r)B''(r)-\frac12B'(r)^2\right].
 \label{eq:gaussian}
\end{equation}
The optical surface element is
\begin{equation}
 \dd S=\sqrt{EG}\,\dd r\dd\phi
 =\frac{r}{B(r)^{3/2}}\dd r\dd\phi.
 \label{eq:surface}
\end{equation}
These two expressions provide the starting point for the Gauss--Bonnet calculation \cite{GibbonsWerner2008,Werner2012}. The full metric contains the terms $-Nr+r^2/\ell^2$. We therefore cannot send the source and observer to a flat infinity unless an outer matching is specified. The cleanest choice is to keep both points at finite distances \cite{RindlerIshak2007,Ishihara2016,OnoIshiharaAsada2017,OnoAsada2019}. Let $u=1/r$ and define
\begin{equation}
 F(u)=\frac{1}{b^2}-u^2B(1/u).
 \label{eq:Fweak}
\end{equation}
A static observer measures the local angle $\Psi$ between the photon direction and the outward radial direction through
\begin{equation}
 \sin\Psi=\frac{b\sqrt{B(r)}}{r}.
 \label{eq:Psi}
\end{equation}
For a source at $r_S$ and a receiver at $r_R$, our sign convention gives
\begin{equation}
 \alpha_{\rm fd}=\Psi_R-\Psi_S+\phi_{RS},
 \label{eq:alphafd}
\end{equation}
with
\begin{equation}
 \phi_{RS}=\int_{u_R}^{u_0}\frac{\dd u}{\sqrt{F(u)}}
 +\int_{u_S}^{u_0}\frac{\dd u}{\sqrt{F(u)}}.
 \label{eq:phiRS}
\end{equation}
The turning point is $u_0$. These equations are exact inside the chosen static region. They keep the quintessence and AdS terms and reduce to the usual infinite-distance angle whenever that limit exists.

\subsection{Local Gauss--Bonnet expansion}
To compare the results with solar-limb bending, we consider a locally matched magnetic-PFDM approximation in which the large-radius terms $-Nr$ and $r^2/\ell^2$ are omitted. The resulting lapse function admits the expansion given in Eq.~\eqref{eq:larger}. The Gauss--Bonnet theorem for the region outside the ray is
\begin{equation}
 \iint_{\mathcal D}\cK\,\dd S+\oint_{\partial\mathcal D}\kappa_g\,\dd t
 +\sum_i\Theta_i=2\pi\chi(\mathcal D).
 \label{eq:GBT}
\end{equation}
The photon path is a geodesic. In the locally flat matching region, the large circular arc gives $\kappa_g\dd t\to\dd\phi$. At leading order, the ray is $r=b/\sin\phi$. Expanding the curvature and surface element gives
\begin{align}
 \cK\dd S={}&\Bigg[-\frac{2M}{r^2}-\frac{3M^2}{r^3}
 +\frac{\lambda}{r^2}\left(\ln\frac{r}{|\lambda|}-\frac32\right)
 +\frac{10MQ_m^3}{r^5}+\cdots\Bigg]\dd r\dd\phi.
 \label{eq:KdS}
\end{align}
The straight-line domain gives the leading mass, PFDM, and magnetic terms. At second post-Minkowskian order, the first correction to the ray boundary must also be included. It changes the incomplete coefficient $3\pi M^2/(4b^2)$ into the correct Schwarzschild value $15\pi M^2/(4b^2)$ \cite{AdlerVirbhadra2022,HeEtAl2020}. The result is
\begin{equation}
 \alpha_{\rm weak}=\frac{4M}{b}+\frac{15\pi M^2}{4b^2}
 -\frac{\lambda}{b}\left[1+2\ln\!\left(\frac{b}{2|\lambda|}\right)\right]
 -\frac{15\pi MQ_m^3}{16b^4}+\cdots 
 \label{eq:alphaweak}
\end{equation}
The PFDM correction has the same $1/b$ order as the leading mass term, but it also contains a logarithm. The first magnetic correction appears only at $b^{-4}$. This is why weak lensing is much more sensitive to the PFDM part of the benchmark path than to the magnetic core.

\subsection{Independent geodesic calculation}
\begin{figure}[ht!]
    \centering
    \begin{minipage}[t]{0.49\textwidth}
        \centering
        \includegraphics[width=\textwidth]{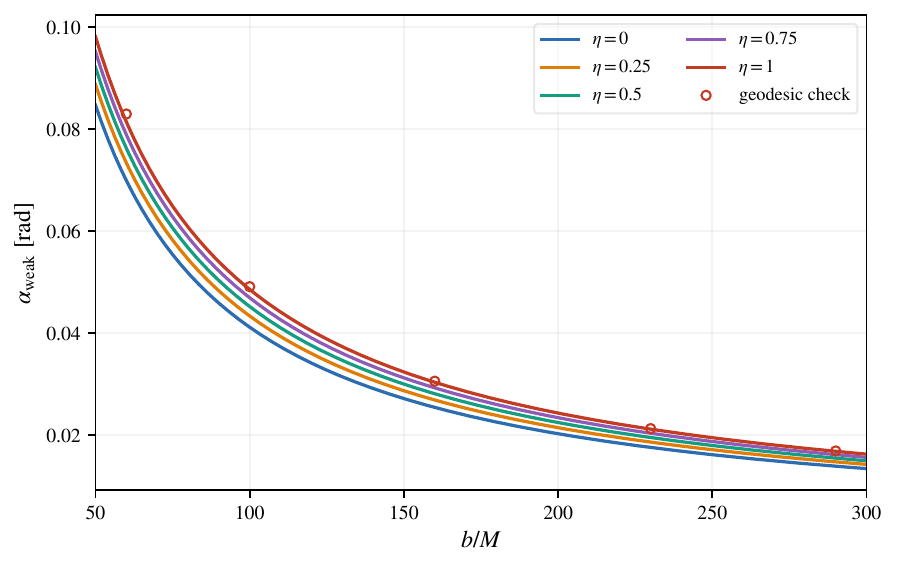}\\
        (a)
    \end{minipage}
    \hfill
    \begin{minipage}[t]{0.49\textwidth}
        \centering
        \includegraphics[width=\textwidth]{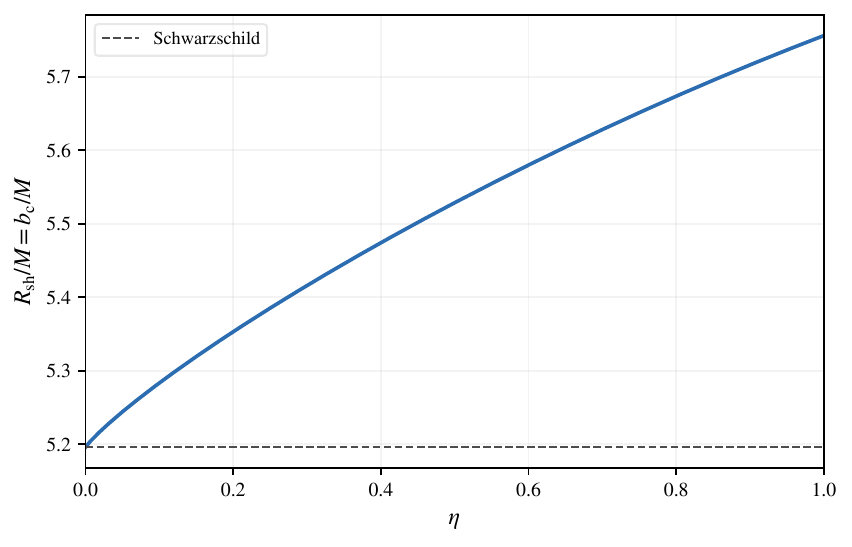}\\
        (b)
    \end{minipage}
    \caption{(a) Weak deflection angle as a function of the impact parameter along the benchmark path. Solid curves show Eq.~\eqref{eq:alphaweak}, while open circles give an independent numerical geodesic check for $\eta=1$. (b) Critical shadow radius along the benchmark path.}
    \label{fig:weak-shadow}
\end{figure}

As an independent check of the weak-deflection angle obtained from the optical geometry, we derive the photon trajectory directly from the null geodesic equations. Using the conserved quantities and radial equation established in Sec.~\ref{sec:geodesics}, Eqs.~\eqref{conserv11}--\eqref{eqmorad2}, the radial equation for the metric in Eq.~\eqref{eq:metric} reads
\begin{equation}
 \dot r^2=E^2-J^2\frac{B(r)}{r^2}.
 \label{eq:nullradial}
\end{equation}
Introducing $u=1/r$ and the impact parameter defined in Eq.~\eqref{eq:impactparameter}, we obtain the orbit equation
\begin{equation}
 \left(\frac{\dd u}{\dd\phi}\right)^2
 =\frac1{b^2}-u^2B(1/u).
 \label{eq:orbit1}
\end{equation}
Differentiating and using the local expansion in Eq.~\eqref{eq:larger}, we find
\begin{equation}
 u''+u=3Mu^2+\frac{\lambda u^2}{2}
 \left[1-3\ln\!\left(\frac{1}{|\lambda|u}\right)\right]
 -6MQ_m^3u^5+\cdots .
 \label{eq:orbit2}
\end{equation}
At first order, we write $u=u_0+u_1$ with $u_0=\sin\phi/b$. The Green-function solution satisfying $u_1(0)=u_1'(0)=0$ is
\begin{equation}
 u_1(\phi)=\int_0^\phi\sin(\phi-\varphi)\,
 \mathcal F[u_0(\varphi)]\,\dd\varphi,
 \label{eq:green}
\end{equation}
where $\mathcal F$ denotes the right-hand side of Eq.~\eqref{eq:orbit2}. To obtain the second mass correction, we replace $3Mu^2$ by $3M(u_0^2+2u_0u_1^{(M)})$. The outgoing condition $u(\pi+\alpha)=0$ then gives Eq.~\eqref{eq:alphaweak}. Appendix~\ref{app:weakderivation} lists the elementary and logarithmic integrals. Fig~\ref{fig:weak-shadow}(a) compares Eq.~\eqref{eq:alphaweak} with a direct integration of the exact magnetic-PFDM geodesic. The relative difference is about $1.9\%$ at $b=60M$ and falls below $0.4\%$ at $b=290M$. The agreement improves as the neglected higher-order terms become smaller.

We can also make a formal solar-limb comparison. We introduce the parametrized post-Newtonian parameter $\gamma_{\rm PPN}$, which measures the amount of spatial curvature produced per unit mass and is equal to unity in General Relativity. The observed angle is written as
\begin{equation}
 \theta=\frac{1+\gamma_{\rm PPN}}{2}\,\delta_{\rm GR},
 \qquad \delta_{\rm GR}=\frac{4M_\odot}{R_\odot},
 \label{eq:ppn}
\end{equation}
with $\gamma_{\rm PPN}=0.99992\pm0.00012$ \cite{Lambert2011}. We then insert the benchmark scaling into Eq.~\eqref{eq:alphaweak} and use $R_\odot/M_\odot\simeq4.7113\times10^5$. This gives
\begin{equation}
 0\leq\eta\lesssim2.06\times10^{-5}
 \qquad (1\sigma,\text{ formal solar analogue}).
 \label{eq:solarbound}
\end{equation}
This number is source dependent. It only tells us that the same local PFDM-magnetic environment cannot be placed unchanged around the Sun. It is not a universal bound on the environment of a supermassive black hole.

\section{Black hole shadow}\label{sec:shadow}
As discussed in Sec.~\ref{sec:geodesics}, the critical impact parameter $b_c$ is determined by the unstable circular photon orbit and sets the critical curve separating captured from scattered photon trajectories. For a static observer at a finite radius $r_o$, an orthonormal tetrad gives \cite{Ladino:2024ned,LADINO2025117031}
\begin{equation}
 \sin\alpha_{\rm sh}=\frac{b_c\sqrt{B(r_o)}}{r_o}.
 \label{eq:finiteangle}
\end{equation}
This is the correct expression inside a non-asymptotically flat static region \cite{PerlickTsupkoBisnovatyi2015,PerlickTsupko2022}. If the local metric is matched to a weak-field exterior, $b_c$ becomes the usual linear critical scale used for angular conversion.

Fig.~\ref{fig:weak-shadow}(b) shows a clear monotonic trend: the critical scale $b_c$ grows along the benchmark path. At $\eta=0$, we recover the Schwarzschild value $3\sqrt3M$. At $\eta=1$, we obtain $b_c\simeq5.7559M$. Table~\ref{tab:strong} also lists the horizon radius, photon-sphere radius, and strong-lensing coefficients. The shadow result is purely geometrical. The brightest ring in an image can lie at another radius because it also depends on the emission model \cite{Gralla2019,Johnson2020,BisnovatyiTsupko2022}. Higher-order photon-ring impact parameters have also been studied as probes of black-hole optical features and their connection with thermodynamic stability \cite{ladino2026naked}. For later use, the critical angular diameter in a matched weak-field region is
\begin{equation}
 d_{\rm crit}=2\theta_\infty
 =\frac{2b_cGM}{c^2D}.
 \label{eq:dcritpre}
\end{equation}
With the central M87* mass and distance values, the diameter $42\,\mu$as corresponds to $\eta\simeq0.444$. Section~\ref{sec:lens} gives the full intervals and explains the difference between a critical diameter and an emission-ring diameter.


\section{Strong deflection by the black hole}\label{sec:strong}

We now turn to rays that pass very close to the unstable photon sphere. Let $r_0$ be the closest approach. In the strong-deflection limit, the bending angle has a universal logarithmic divergence plus a finite part \cite{Darwin1959,BozzaEtAl2001,Bozza2002,Bozza2010,Eiroa2012,Tsukamoto2018}. At the turning point,
\begin{equation}
 A(r_0)\dot t_0^2=C(r_0)\dot\phi_0^2,
 \label{eq:turning}
\end{equation}
so
\begin{equation}
 b(r_0)=\sqrt{\frac{C(r_0)}{A(r_0)}}.
 \label{eq:br0}
\end{equation}

For our metric, the exact coordinate-angle integral from $r_0$ to a matching radius $r_{\rm max}$ is
\begin{equation}
 I(r_0)=2\int_{r_0}^{r_{\rm max}}
 \frac{b(r_0)\,\dd r}{r^2
 \sqrt{1-b(r_0)^2B(r)/r^2}},
 \qquad \alpha=I-\pi.
 \label{eq:deflectionI}
\end{equation}

For an asymptotically flat metric, one takes $r_{\rm max}\to\infty$. In the present case, however, the non-asymptotically flat geometry requires a finite matching surface. Although the divergent coefficient is determined by the photon-sphere region, the finite contribution retains a mild dependence on the chosen matching radius \cite{BozzaScarpetta2007}. To analyze these contributions, we next introduce a regularization variable that maps the radial integration domain to a finite interval.

\subsection{Regularization with $z=1-r_0/r$}

The variable $z$ maps the integration interval $r\in[r_0,r_{\rm max}]$ to $z\in[0,z_{\rm max}]$, facilitating the isolation of the logarithmic divergence as the closest-approach radius approaches the photon-sphere radius. We define
\begin{equation}
 z=1-\frac{r_0}{r},\qquad r=\frac{r_0}{1-z},
 \qquad z_{\rm max}=1-\frac{r_0}{r_{\rm max}}.
 \label{eq:zstrong}
\end{equation}
\begin{figure}[ht!]
    \centering
    \begin{minipage}[t]{0.49\textwidth}
        \centering
        \includegraphics[width=\textwidth]{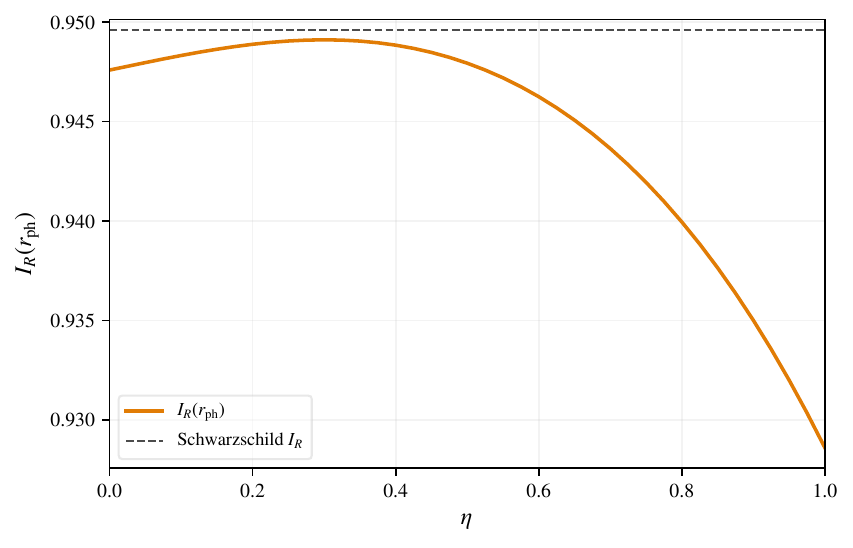}\\
        (a)
    \end{minipage}
    \hfill
    \begin{minipage}[t]{0.49\textwidth}
        \centering
        \includegraphics[width=\textwidth]{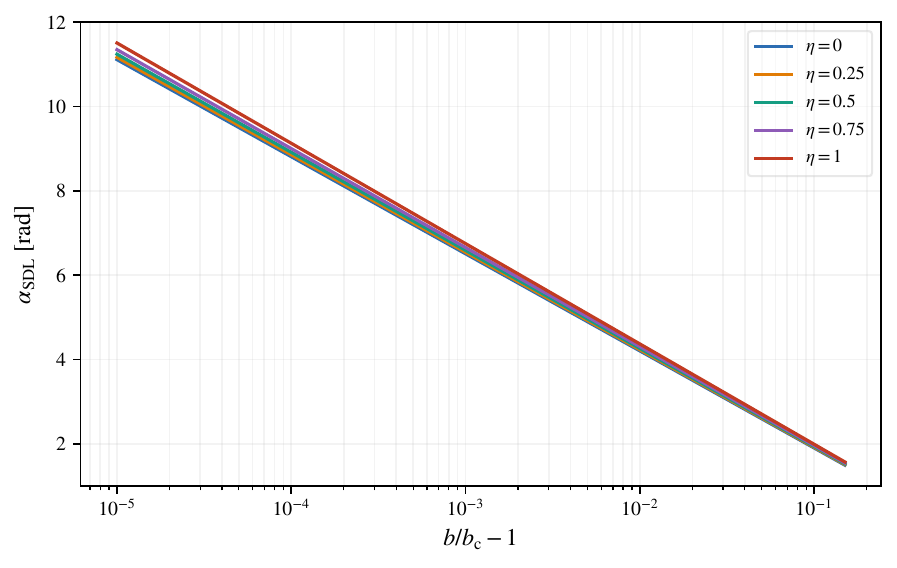}\\
        (b)
    \end{minipage}
    \caption{(a) Regular strong-deflection contribution $I_R(r_{\rm ph})$ along the benchmark path for the adopted matching radius. The dashed line indicates the asymptotic Schwarzschild value $0.9496$. (b) Strong-deflection profile as a function of $b/b_c-1$, showing the logarithmic divergence for every value of $\eta$.}
    \label{fig:strong-deflection}
\end{figure}
Furthermore, since $b^2=r_0^2/B_0$, where $B_0=B(r_0)$, the integral becomes
\begin{equation}
 I(r_0)=2\int_0^{z_{\rm max}}
 \frac{\dd z}{\sqrt{G(z,r_0)}},
 \label{eq:Iz}
\end{equation}
with
\begin{equation}
 G(z,r_0)=B_0-B\!\left(\frac{r_0}{1-z}\right)(1-z)^2.
 \label{eq:Gstrong}
\end{equation}
Close to $z=0$,
\begin{equation}
 G(z,r_0)=p(r_0)z+q(r_0)z^2+O(z^3),
 \label{eq:Gexpand}
\end{equation}
where
\begin{align}
 p(r_0)&=2B_0-r_0B'_0,\label{eq:pstrong}\\
 q(r_0)&=-B_0+r_0B'_0-\frac12r_0^2B''_0.\label{eq:qstrong}
\end{align}
The photon-sphere equation is simply $p(r_{\rm ph})=0$. At the photon sphere,
\begin{equation}
 q_m=B_m-\frac12r_{\rm ph}^2B_m'',
 \label{eq:qm}
\end{equation}

where the subscript $m$ denotes evaluation at $r_{\rm ph}$. For an unstable orbit, $q_m>0$. To separate the finite part, we subtract the leading singular model:
\begin{align}
 I_R(r_0;r_{\rm max})=2\int_0^{z_{\rm max}}\Bigg[
 \frac{1}{\sqrt{G(z,r_0)}}
 -\frac{1}{\sqrt{p(r_0)z+q(r_0)z^2}}
 \Bigg]\dd z.
 \label{eq:IRdef}
\end{align}

The subtracted integral is regular, while the model integral can be evaluated analytically. As $r_0\to r_{\rm ph}$, the impact parameter approaches its critical value $b_c$ quadratically, and the bending angle exhibits a logarithmic divergence of the form
\begin{equation}
 \alpha(b)=-\bar a\ln\!\left(\frac{b}{b_c}-1\right)+\bar b
 +O[(b-b_c)\ln(b-b_c)].
 \label{eq:sdl}
\end{equation}
Here, $\bar a$ determines the strength of the logarithmic divergence and is fixed by the metric near the photon sphere, whereas $\bar b$ represents the finite contribution, which depends on the near-photon-sphere geometry and the remainder of the trajectory up to the matching radius $r_{\rm max}$. For the metric in Eq.~\eqref{eq:metric}, we obtain
\begin{equation}
\bar a=\frac{1}{\sqrt{q_m}}
 =\left[\frac{2}{2B_m-r_{\rm ph}^2B_m''}\right]^{1/2}.
 \label{eq:abar}
\end{equation}
With the standard asymptotically flat normalization,
\begin{equation}
\bar b=-\pi+I_R(r_{\rm ph})
 +\bar a\ln\!\left(\frac{2q_m}{B_m}\right).
 \label{eq:bbar}
\end{equation}
For the locally matched geometry, we use the same expression but evaluate $I_R$ at the chosen $r_{\rm max}$. Appendix~\ref{app:strongderivation} shows how the quadratic minimum of $b(r_0)$ produces the logarithm in Eq.~\eqref{eq:sdl}. Moreover, for $\eta=0$ and $r_{\rm max}=5000M$, we obtain $I_R\simeq0.9476$ and $\bar b\simeq-0.4022$. Increasing the matching radius to $20000M$ gives $I_R\simeq0.9491$ and $\bar b\simeq-0.4007$. These values approach the standard Schwarzschild limits, $0.9496$ and $-0.4002$. Appendix~\ref{app:numerics} gives the convergence details.

Fig.~\ref{fig:strong-deflection}(a) shows the regular contribution $I_R(r_{\rm ph})$ along the benchmark path. It varies with $\eta$ and remains close to the asymptotic Schwarzschild value indicated by the dashed line. Fig.~\ref{fig:strong-deflection}(b) displays the strong-deflection profile as a function of $b/b_c-1$, illustrating the logarithmic divergence as $b\to b_c^+$ for all considered values of $\eta$.
Finally, Table~\ref{tab:strong} summarizes the horizon radius, photon-sphere radius, critical scale, and strong-deflection coefficients. From $\eta=0$ to $\eta=1$, the critical scale grows by about $10.8\%$, while the change in $\bar a$ is smaller. The coefficient $\bar b$ remains close to its Schwarzschild value because the change in the regular integral partly cancels the change in the logarithmic term.

\begin{table}[tbp]
 \centering
 \caption{Geometrical and strong-lensing quantities along the benchmark path.}
 \label{tab:strong}
 \begin{tabular}{c c c c c c c}
\toprule
$\eta$ & $x_h$ & $x_{\rm ph}$ & $b_c/M$ & $\bar a$ & $\bar b$ & $I_R$\\
\midrule
0.00 & 2.0000 & 3.0000 & 5.1962 & 1.0000 & $-0.4022$ & 0.9476\\
0.25 & 2.0650 & 3.0988 & 5.3852 & 1.0042 & $-0.3995$ & 0.9479\\
0.75 & 2.1393 & 3.2233 & 5.6512 & 1.0192 & $-0.3875$ & 0.9419\\
1.00 & 2.1541 & 3.2620 & 5.7559 & 1.0332 & $-0.3908$ & 0.9286\\
\bottomrule
\end{tabular}
\end{table}

\begin{figure}[ht!]
 \centering
 \includegraphics[width=0.94\textwidth]{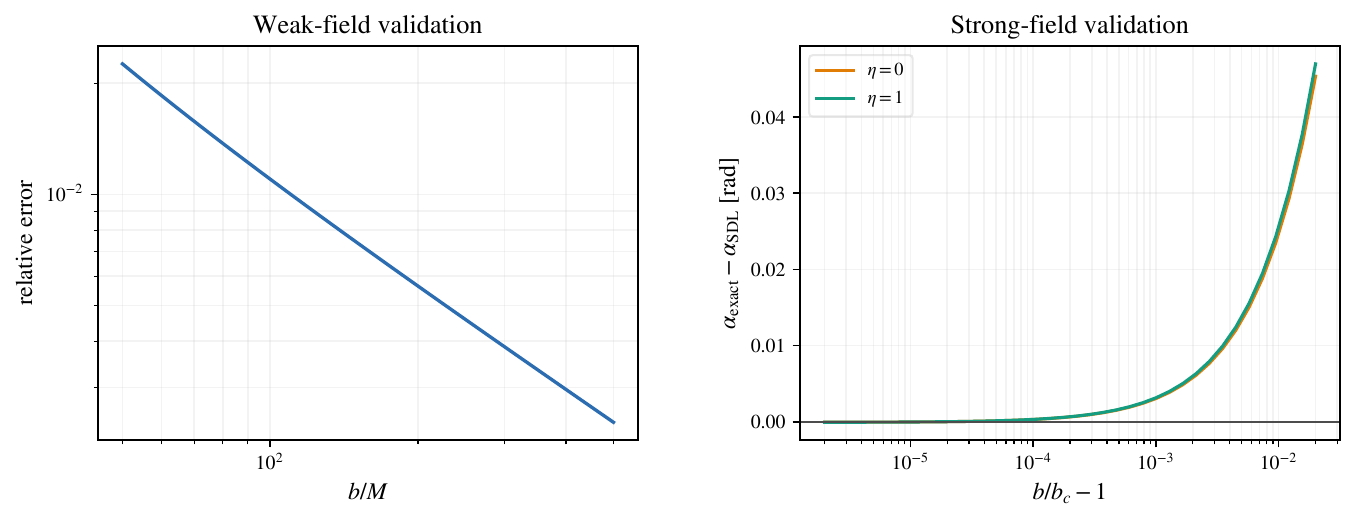}
 \caption{Numerical validation. Left: relative difference between the analytic weak-field expansion and the exact magnetic-PFDM geodesic for $\eta=1$. Right: residual of the strong-deflection logarithmic approximation for Schwarzschild and the full benchmark.}
 \label{fig:validation}
\end{figure}

Fig.~\ref{fig:validation} shows the range of validity of both approximations. The left panel compares the analytic weak-field angle with the exact geodesic result. The right panel shows the difference between the exact angle and the logarithmic strong-deflection formula. The residual tends to zero as $b\to b_c^+$.

\section{Lens equation, relativistic images, and observational intervals}\label{sec:lens}
The strong-deflection regime allows us to investigate relativistic images formed by photons winding around the black hole. Their positions and angular separations depend on the critical impact parameter and the strong-deflection coefficients. To connect these quantities with observable image positions, we consider the nearly aligned lensing geometry.

\subsection{Relativistic images in the nearly aligned geometry}

\begin{figure}[tbp]
 \centering
 \includegraphics[width=0.96\textwidth]{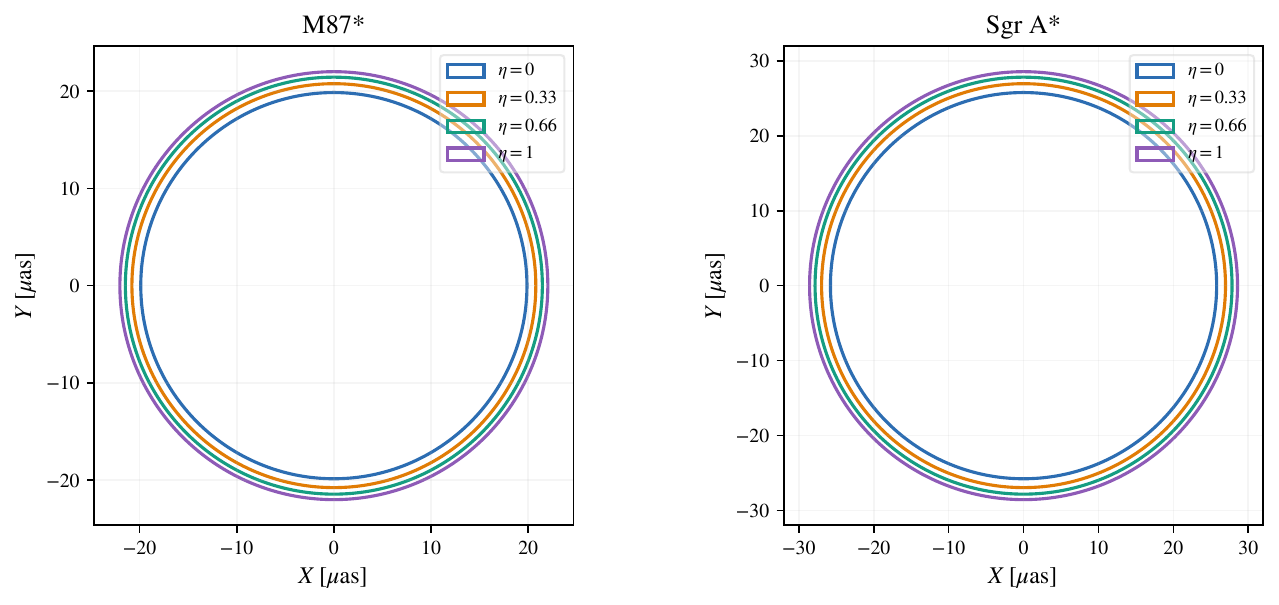}
 \caption{Outermost relativistic Einstein rings for M87* and Sgr~A* along the benchmark path. The circles show the geometrical strong-lensing scale, not a simulated emission image.}
 \label{fig:rings}
\end{figure}

We use the standard thin-lens notation. The distances $D_{OL}$, $D_{LS}$, and $D_{OS}$ connect the observer, lens, and source. The source angle is $\psi$, and the image angle is $\theta$. For a nearly aligned source, the small-angle lens equation is \cite{VirbhadraEllis2000,BozzaEtAl2001,Bozza2002,BozzaMancini2004}

\begin{equation}
 \psi=\theta-\frac{D_{LS}}{D_{OS}}\Delta\alpha_n,
 \qquad
 \Delta\alpha_n=\alpha-2n\pi.
 \label{eq:lenseq}
\end{equation}
We insert the strong-deflection expression and define $\theta_n^0$ by $\alpha(\theta_n^0)=2n\pi$. This gives
\begin{equation}
 \theta_n^0=\frac{b_c}{D_{OL}}(1+\eps_n),
 \qquad
 \eps_n=\exp\!\left(\frac{\bar b-2n\pi}{\bar a}\right).
 \label{eq:theta0}
\end{equation}
The small factor $\eps_n$ shows how quickly the relativistic images accumulate near the limiting angle. Expanding $\alpha(D_{OL}\theta)$ around $\theta_n^0$ gives
\begin{equation}
 \Delta\alpha_n=-\frac{\bar aD_{OL}}{b_c\eps_n}\Delta\theta_n.
 \label{eq:dalpha}
\end{equation}
The lens equation then gives
\begin{equation}
 \theta_n=\theta_n^0+
 \frac{b_c\eps_nD_{OS}}{\bar aD_{LS}D_{OL}}
 (\psi-\theta_n^0).
 \label{eq:thetan}
\end{equation}
For exact alignment, $\psi=0$, the two images join into a relativistic Einstein ring. To leading order,
\begin{equation}
 \theta_n^E\simeq\frac{b_c}{D_{OL}}(1+\eps_n).
 \label{eq:einstein}
\end{equation}
The extra distance-ratio correction is proportional to $(b_c/D_{OL})\eps_n$ and is negligible for the angular scales used here. Furthermore, Fig.~\ref{fig:rings} shows the outermost ring, $n=1$, for M87* and Sgr~A*. We use the same central mass and distance values as Ref.~\cite{OvgunFathi2025}. The circles show only the geometrical strong-lensing scale; they are not simulated plasma images.

The signed magnification is the inverse Jacobian of the lens map. At leading order,
\begin{equation}
 \mu_n=\left|\frac{\psi}{\theta}\frac{\dd\psi}{\dd\theta}\right|^{-1}_{\theta_n^0}
 =\frac{b_c^2\eps_n(1+\eps_n)D_{OS}}
 {\bar a\,|\psi|D_{LS}D_{OL}^2}.
 \label{eq:magnification}
\end{equation}
The limiting angle is
\begin{equation}
 \theta_\infty=\lim_{n\to\infty}\theta_n^0=\frac{b_c}{D_{OL}}.
 \label{eq:thetainf}
\end{equation}
If the first image can be separated from the packed inner images, two useful observables are \cite{Bozza2002,BozzaMancini2004,Bozza2010}
\begin{align}
 s&=\theta_1-\theta_\infty\simeq\theta_\infty\eps_1,
 \label{eq:ssep}\\\
 r_{\rm mag}&=2.5\log_{10}\!\left[\exp\!\left(\frac{2\pi}{\bar a}\right)\right].
 \label{eq:rmag}
\end{align}
The first is an angular separation. The second is the magnitude difference between the first image and the remaining sequence in the standard strong-deflection approximation. Table~\ref{tab:rings} lists both. The relative separation is only of order $10^{-3}$, so resolving the sequence directly would be very difficult.

\begin{table}[tbp]
 \centering
 \caption{Outermost relativistic ring and strong-lensing observables. Angular radii are in $\mu$as.}
 \label{tab:rings}
 \begin{tabular}{c c c c c}
\toprule
$\eta$ & $\theta_1^{E}$(M87*) & $\theta_1^{E}$(Sgr A*) & $(s/\theta_\infty)\times 10^3$ & $r_{\rm mag}$\\
\midrule
0.00 & 19.869 & 25.774 & 1.249 & 6.822\\
0.25 & 20.593 & 26.713 & 1.293 & 6.794\\
0.50 & 21.143 & 27.426 & 1.352 & 6.754\\
0.75 & 21.613 & 28.036 & 1.437 & 6.693\\
1.00 & 22.017 & 28.560 & 1.565 & 6.603\\
\bottomrule
\end{tabular}
\end{table}

\subsection{Angular conversion and simple EHT intervals}

For a matched weak-field exterior, the critical angular diameter is
\begin{equation}
 2\theta_\infty=\frac{2b_cGM}{c^2D}.
 \label{eq:angular}
\end{equation}
It is useful to define
\begin{equation}
 \theta_g=\frac{GM}{c^2D},
 \qquad 2\theta_\infty=2\frac{b_c}{M}\theta_g.
 \label{eq:thetag}
\end{equation}
If the dimensionless ratio $b_c/M$ is fixed, the mass and distance errors give
\begin{equation}
 \left(\frac{\sigma_{2\theta}}{2\theta_\infty}\right)^2
 =\left(\frac{\sigma_M}{M}\right)^2
 +\left(\frac{\sigma_D}{D}\right)^2
 -2\rho_{MD}\frac{\sigma_M\sigma_D}{MD},
 \label{eq:errorprop}
\end{equation}
where $\rho_{MD}$ is the mass-distance correlation. We do not run a full correlated mass-distance analysis here. We include this formula so that the numerical intervals can be updated for any chosen external prior.

For M87*, we use the observed ring diameter $42\pm3\,\mu$as \cite{EHT2019I,EHT2019IV,EHT2019VI}. For Sgr~A*, we adopt the conservative $1\sigma$ interval $41.7$--$55.6\,\mu$as, which covers the range of EHT reconstructions and model calibrations \cite{EHT2022I,EHT2022III,EHT2022IV,EHT2022V}. An observed bright ring is not identical to a vacuum critical curve. The comparison below is therefore a simple geometrical size test, not a likelihood fit \cite{Gralla2019,PerlickTsupko2022,VagnozziEtAl2023}.

\begin{figure}[tbp]
 \centering
 \includegraphics[width=0.96\textwidth]{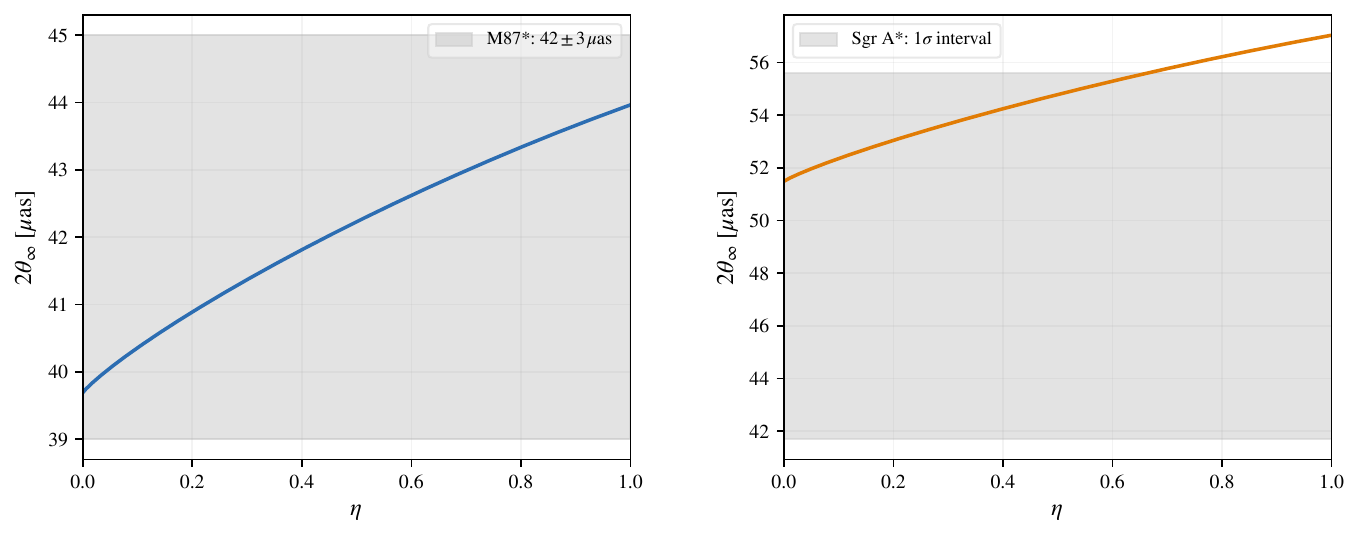}
 \caption{Theoretical critical angular diameter along the benchmark path. The shaded regions show the adopted $1\sigma$ intervals for M87* and Sgr~A*.}
 \label{fig:eht}
\end{figure}

Fig.~\ref{fig:eht} compares the theoretical critical angular diameter along the benchmark path with the adopted observational intervals for M87* and Sgr~A*. The overlap between the theoretical curve and these bands provides a simple geometrical test of the benchmark parameter $\eta$.

With the central mass and distance values used here, the M87* band contains the full interval $0\leq\eta\leq1$. The upper edge of the Sgr~A* band gives
\begin{equation}
 0\leq\eta\lesssim0.665.
 \label{eq:sgrbound}
\end{equation}
These are one-dimensional intervals along the benchmark path. They are not separate constraints on $Q_m$, $\lambda$, $N$, and $\ell$. A realistic inference would need rotation, a global matching prescription, nuisance parameters, scattering for Sgr~A*, and calibrated interferometric data \cite{Chael2016,Chael2018,BlackburnEtAl2020,EHT2022II,EHT2022V}.

\begin{table}[tbp]
 \centering
 \caption{Formal weak-field and EHT benchmark-path constraints. The Solar-System line is source dependent and must not be combined directly with the black hole environment bounds.}
 \label{tab:constraints}
\begin{tabular}{l  c  c}
\toprule
Observable & Adopted interval & Benchmark-path constraint\\
\midrule
Solar-limb PPN & \quad$\gamma_{\rm PPN}=0.99992\pm0.00012$ & \quad $0\leq\eta\lesssim 2.06\times 10^{-5}$ (formal local bound)\\
M87* diameter & $39$--$45\,\mu$as & $0\leq\eta\leq 1.000$\\
Sgr A* diameter & $41.7$--$55.6\,\mu$as & $0\leq\eta\lesssim 0.665$\\
\bottomrule
\end{tabular}
\end{table}
\section{Discussion}\label{sec:discussion}

The results give a simple physical picture. Far from the black hole, the PFDM term is the most visible correction because it enters the weak angle through a logarithmic $1/b$ term. The magnetic contribution first appears at $b^{-4}$ and is strongly suppressed. Close to the photon sphere, this suppression is gone. The magnetic core, PFDM, quintessence, and AdS curvature then act together on the horizon, photon sphere, and strong-deflection coefficients. Similar differences between environmental effects and nonlinear-electromagnetic effects have been found in related models \cite{PantigOvgun2022,Uniyal2023,OkyayOvgun2022,SuQiao2025,WaseemEtAl2025}.

The Jacobi calculation gives the expected local temperature. For a simple horizon, the Hawking temperature depends only on the first derivative of the lapse. The particle mass and all regular terms in the radial momentum disappear from the pole residue. This agrees with Hamilton--Jacobi and null-geodesic tunnelling methods \cite{ParikhWilczek2000,Angheben2005,VanzoAcquavivaDiCriscienzo2011,BeraGhoshMajhi2020}. Along our path, the horizon grows while the dimensionless surface gravity decreases. We also calculate the Hawking sparsity and find a particularly strong dependence on the quintessence parameter $N$. Increasing $N$ can induce a non-monotonic behavior, with the sparsity increasing for intermediate-size black holes before decreasing at larger horizon radii. Although this effect is physically significant, its direct astrophysical observability remains challenging.

Weak lensing requires the most care because the full metric is not asymptotically flat. We therefore wrote the exact finite-distance angle first. For the solar comparison, we used a local matched slice and did not extend the quintessence and AdS terms to infinity. We also kept the full second-post-Minkowskian mass coefficient, $15\pi/4$. This coefficient is missed if one integrates only over a straight-line domain. The exact geodesic check in Fig.~\ref{fig:validation} shows the remaining truncation error. This is the same logic used in finite-distance lensing for asymptotically curved spacetimes \cite{RindlerIshak2007,Ishihara2016,OnoIshiharaAsada2017,OnoAsada2019}.

The photon sphere is the local extremum of $B/r^2$, and its critical impact parameter sets the shadow scale. Along the benchmark path, $r_h/M$, $r_{\rm ph}/M$, and $b_c/M$ all increase. Still, they do not grow by the same fraction. The coefficient $\bar a$ changes by only a few percent because it depends on the curvature of the photon potential, while $b_c$ changes by about eleven percent. This shows why the shadow size and the strong-lensing coefficients carry different information \cite{CunhaHerdeiro2018,PerlickTsupko2022,PantigOvgun2025}.

The regularization also separates the local and nonlocal strong-lensing quantities. The coefficient $\bar a$ depends only on the photon-sphere region. The coefficient $\bar b$ contains the regular integral $I_R$ and therefore keeps a small memory of the outer matching radius. Appendix~\ref{app:numerics} shows that the Schwarzschild values approach their standard limits as $r_{\rm max}$ is increased. For our full benchmark, the same convergence is found. The matching effect is much smaller than the change in $b_c$ along the path.

The EHT comparison must also be read with care. A vacuum critical diameter is not the same as the diameter of a bright plasma ring after radiative transfer and beam convolution \cite{Gralla2019,Johnson2020,BronzwaerFalcke2021}. M87* and Sgr~A* also rotate, and Sgr~A* is variable and affected by interstellar scattering \cite{EHT2019V,EHT2022II,EHT2022III,EHT2022IV}. For these reasons, we report only simple benchmark-path intervals. We do not interpret them as separate measurements of the four model parameters.

Within this limited scope, the full calculation is reproducible. The appendices list the derivatives, the contour choice, the weak-field integrals, the strong-deflection subtraction, the angular conversion, the root selection, and the convergence tests. The supplied code uses standard Python tools \cite{Hunter2007,Harris2020,Virtanen2020} and regenerates every figure and table.

\section{Conclusion}\label{sec:conclusion}

We have investigated the thermal and optical properties of a magnetic AdS black hole surrounded by PFDM and quintessence. Starting from the spacetime geometry, we examined its curvature structure and Hawking radiation, and then explored its gravitational lensing properties across the weak- and strong-deflection regimes. We further analyzed the photon sphere, black hole shadow, relativistic images, and the compatibility of the predicted shadow sizes with selected EHT intervals. This sequence provides a unified picture of how the surrounding matter and magnetic charge affect different observable and thermodynamic properties of the spacetime, while detailed derivations and numerical consistency checks are provided in the appendices.

For the quintessence state parameter $w=-2/3$, the curvature and gradient invariants characterize the geometric structure and help identify the horizon locations. The Ricci scalar separates the contributions associated with the magnetic charge, PFDM, quintessence, and the AdS curvature, confirming that the resulting spacetime is not simply a rescaled Schwarzschild geometry. The Jacobi-metric approach to Hawking radiation also recovers the standard relation between the Hawking temperature and the radial derivative of the metric function at the event horizon. In this calculation, the mass of the emitted particle does not affect the pole residue.

Following this temperature calculation, we examined the sparsity of Hawking radiation to connect the thermal properties of the black hole with its optical structure. The emission remains highly sparse, with a Schwarzschild benchmark value of $\mathcal{S}_{\rm Sch}\simeq73.5$. Along the selected benchmark path, the sparsity increases to $\mathcal{S}=83.56$. More generally, it increases with the magnetic charge, the magnitude of the PFDM parameter, and the quintessence intensity, while decreasing as the horizon radius grows. The termination of the sparsity branches is associated with the disappearance of the unstable photon sphere, revealing a connection between the relevant optical structure and the behavior of thermal emission. Although direct measurements of Hawking sparsity remain challenging for astrophysical black holes, combining shadow observations with strong-lensing measurements may provide complementary tests of departures from the Schwarzschild geometry. 

Building on these geometric results, we investigated gravitational bending using both exact finite-distance equations and a locally matched weak-deflection expansion. The former retains the full non-asymptotically flat geometry, whereas the latter provides a transparent description of the leading corrections to the Schwarzschild bending angle. The two approaches agree increasingly well as the impact parameter grows. Applying the same local benchmark environment to a formal solar-limb comparison yields $\eta\lesssim2.06\times10^{-5}$. This estimate should not, however, be interpreted as a universal constraint on black holes, since PFDM and quintessence describe the environment surrounding the source and the result depends on the adopted benchmark assumptions.

The weak-deflection analysis naturally leads to the study of the photon sphere and shadow, which encode the behavior of null geodesics closer to the black hole. Along the selected benchmark path, the outer horizon increases from $2M$ to $2.1541M$, while the photon-sphere radius grows from $3M$ to $3.2620M$. Accordingly, the critical impact parameter increases from $5.1962M$ to $5.7559M$. These changes demonstrate that the environmental and magnetic contributions modify not only the horizon structure but also the characteristic scales governing the capture of light. In the strong-deflection regime, the logarithmic coefficient increases from its Schwarzschild value of $\bar a=1$ to approximately $\bar a=1.0332$. The regular coefficient was obtained by subtracting the analytic logarithmic contribution, and its numerical stability was checked against variations of the matching radius.

These optical modifications can then be compared with current EHT shadow-size intervals. Using the central mass and distance values adopted in this work, the M87* interval accommodates the full benchmark path, whereas the conservative Sgr~A* interval yields $\eta\lesssim0.665$. These comparisons provide geometrical consistency intervals rather than a simultaneous fit of the model parameters to EHT observations. A more complete observational analysis would require a rotating spacetime, a globally consistent exterior matching, and calibrated visibility-domain data.

Taken together, these results show how magnetic charge, PFDM, and quintessence affect the thermal and optical properties of the black hole. The framework developed here provides a systematic and reproducible basis for investigating these effects. Future work could combine independent astrophysical observables and compare alternative black hole geometries to identify distinctive signatures and further constrain the model parameters.

\medskip
\noindent\textbf{Acknowledgements.}
M.F. acknowledges financial support from the Agencia Nacional de Investigaci\'{o}n y Desarrollo (ANID), Chile, through FONDECYT Postdoctoral project No.~3260029. C.R.F. acknowledge support from Conahcyt-Mexico, grant No. 4003366. J.R.V. is partially supported by the Centro de F\'{i}sica Te\'{o}rica de Valpara\'{i}so (CeFiTeV).

\medskip
\noindent\textbf{Declaration of competing interest.}
The authors declare that they have no known competing financial interests or personal relationships that could have influenced the work reported in this paper.

\medskip
\noindent\textbf{Data and code availability.}
All figures and tables are generated with NumPy, SciPy, and Matplotlib \cite{Hunter2007,Harris2020,Virtanen2020}. The script saves the numerical results in JSON format and writes the LaTeX table files automatically. No random sampling is used, so the results are deterministic up to the stated numerical tolerances. The numerical scripts and figure data are available from the corresponding author upon reasonable request.

\appendix

\section{Derivatives and curvature identities}\label{app:derivatives}

For $w=-2/3$, the lapse is
\begin{equation}
 B(r)=1-\frac{2Mr^2}{r^3+Q_m^3}
 +\frac{\lambda}{r}\ln\!\left(\frac{r}{|\lambda|}\right)
 -Nr+\frac{r^2}{\ell^2}.
\end{equation}
The first three derivatives used in the paper are
\begin{align}
 B'(r)={}&\frac{2Mr(r^3-2Q_m^3)}{(r^3+Q_m^3)^2}
 +\frac{\lambda}{r^2}\left[1-\ln\!\left(\frac{r}{|\lambda|}\right)\right]
 -N+\frac{2r}{\ell^2},
 \label{eq:Bprime}\\
 B''(r)={}&-\frac{4M(r^6-7Q_m^3r^3+Q_m^6)}{(r^3+Q_m^3)^3}
 +\frac{\lambda}{r^3}\left[2\ln\!\left(\frac{r}{|\lambda|}\right)-3\right]
 +\frac{2}{\ell^2},
 \label{eq:Bsecond}\\
 B'''(r)={}&\frac{12Mr^2(r^6-16Q_m^3r^3+10Q_m^6)}{(r^3+Q_m^3)^4}
 +\frac{\lambda}{r^4}\left[11-6\ln\!\left(\frac{r}{|\lambda|}\right)\right].
 \label{eq:Bthird}
\end{align}
For $w=-2/3$, the quintessence term contributes to $B'$ but not to the higher derivatives. The AdS term contributes to $B'$ and $B''$, but not to $B'''$.

For reference, the nonzero orthonormal Riemann components may be written as
\begin{equation}
 R_{\hat t\hat r\hat t\hat r}=-\frac12B'',\qquad
 R_{\hat t\hat\theta\hat t\hat\theta}
 =R_{\hat t\hat\phi\hat t\hat\phi}=-\frac{B'}{2r},
\end{equation}
\begin{equation}
 R_{\hat r\hat\theta\hat r\hat\theta}
 =R_{\hat r\hat\phi\hat r\hat\phi}=\frac{B'}{2r},
 \qquad
 R_{\hat\theta\hat\phi\hat\theta\hat\phi}=\frac{1-B}{r^2}.
\end{equation}
Squaring these components with the correct multiplicities gives Eq.~\eqref{eq:kretsch}. Their contraction gives Eq.~\eqref{eq:ricci}. Substituting Eqs.~\eqref{eq:Bprime} and \eqref{eq:Bsecond} then gives Eq.~\eqref{eq:ricciexplicit}.

\section{Jacobi reduction and contour prescription}\label{app:jacobi}

For a static metric $\dd s^2=-V^2\dd t^2+g_{ij}\dd x^i\dd x^j$, the separated Hamilton--Jacobi equation is
\begin{equation}
 -\frac{E^2}{V^2}+g^{ij}\partial_iS_J\partial_jS_J+m^2=0.
\end{equation}
It may be rewritten as
\begin{equation}
 j^{ij}\partial_iS_J\partial_jS_J=1,
 \qquad
 j_{ij}=\frac{E^2-m^2V^2}{V^2}g_{ij}.
\end{equation}
For $V^2=B$ and $g_{rr}=1/B$, this is Eq.~\eqref{eq:jacobi}. The radial action is
\begin{equation}
 S_J^{(r)}=\mp\int\frac{\sqrt{E^2-m^2B(r)}}{B(r)}\dd r.
\end{equation}
Near a simple horizon, set $r-r_h=\epsilon e^{i\vartheta}$. Choosing the contour above or below the pole only changes the sign convention. The singular part is
\begin{equation}
 \int\frac{\dd r}{r-r_h}=\pm i\pi,
\end{equation}
while the regular terms vanish as $\epsilon\to0$. This gives Eq.~\eqref{eq:imS}. The derivation assumes a simple horizon. An extremal double root has a different pole structure and must be treated separately \cite{Angheben2005,VanzoAcquavivaDiCriscienzo2011}.

\section{Weak-field algebra and finite-distance limit}\label{app:weakderivation}

The straight-line Gauss--Bonnet calculation uses the radial integrals
\begin{align}
 \int_R^\infty\frac{\dd r}{r^2}&=\frac1R,\\\
 \int_R^\infty\frac{\ln(r/a)-3/2}{r^2}\dd r
 &=\frac{\ln(R/a)-1/2}{R},\\\
 \int_R^\infty\frac{\dd r}{r^5}&=\frac{1}{4R^4}.
\end{align}
With $R=b/\sin\phi$, the angular integrals are
\begin{equation}
 \int_0^\pi\sin\phi\,\dd\phi=2,
 \qquad
 \int_0^\pi\sin\phi\ln(\sin\phi)\dd\phi=2\ln2-2,
\end{equation}
\begin{equation}
 \int_0^\pi\sin^2\phi\,\dd\phi=\frac\pi2,
 \qquad
 \int_0^\pi\sin^4\phi\,\dd\phi=\frac{3\pi}{8}.
\end{equation}
These relations give the leading mass, PFDM, magnetic, and straight-line $M^2$ terms. The straight-line domain alone gives $3\pi M^2/(4b^2)$. At second post-Minkowskian order, the first displacement of the ray boundary adds $3\pi M^2/b^2$, producing the complete coefficient $15\pi/4$.

The geodesic method shows the same correction directly. For the mass part,
\begin{equation}
 u_0=\frac{\sin\phi}{b},\qquad
 u_1^{(M)}=\frac{M}{b^2}(1+\cos^2\phi)
\end{equation}
is a symmetric first-order solution. At the next order, the source contains $6Mu_0u_1^{(M)}$. Solving with the Green kernel in Eq.~\eqref{eq:green} and expanding the outgoing zero gives $15\pi M^2/(4b^2)$. For the PFDM part, we separate $\ln(|\lambda|u_0)^{-1}$ into a constant term and $\ln(\sin\phi)$. The logarithmic integral written above then gives the PFDM term in Eq.~\eqref{eq:alphaweak}.

For the full finite-distance geometry, a weak-field expansion is not needed. We first solve $F(u_0)=0$ for the turning point. We then evaluate the two integrals in Eq.~\eqref{eq:phiRS} and calculate the endpoint angles from Eq.~\eqref{eq:Psi}. This procedure keeps both $N$ and $\ell$ and can be used for any specified source and observer positions.

\section{Derivation of the strong-deflection logarithm}\label{app:strongderivation}

For $b(r)=r/\sqrt{B(r)}$, the photon-sphere condition implies $b'(r_{\rm ph})=0$. The second logarithmic derivative is
\begin{equation}
 \left.\frac{\dd^2\ln b}{\dd r^2}\right|_{r_{\rm ph}}
 =\frac{q_m}{B_mr_{\rm ph}^2}.
\end{equation}
Therefore,
\begin{equation}
 \frac{b}{b_c}-1
 =\frac{q_m}{2B_mr_{\rm ph}^2}(r_0-r_{\rm ph})^2
 +O[(r_0-r_{\rm ph})^3].
 \label{eq:bquad}
\end{equation}
The singular model integral is
\begin{equation}
 I_D=2\int_0^{z_{\rm max}}\frac{\dd z}{\sqrt{pz+qz^2}}
 =\frac{4}{\sqrt q}\operatorname{arsinh}\sqrt{\frac{qz_{\rm max}}{p}}.
\end{equation}
As $p\to0^+$,
\begin{equation}
 I_D=-\frac{2}{\sqrt q}\ln p
 +\frac{4}{\sqrt q}\ln(2\sqrt{qz_{\rm max}})+O(p\ln p).
\end{equation}
Moreover,
\begin{equation}
 p(r_0)=\frac{2q_m}{r_{\rm ph}}(r_0-r_{\rm ph})+O[(r_0-r_{\rm ph})^2].
\end{equation}
Combining this relation with Eq.~\eqref{eq:bquad} converts $\ln p$ into one half of $\ln(b/b_c-1)$ and gives $\bar a=1/\sqrt{q_m}$. The remaining constants combine with the regular integral to give Eq.~\eqref{eq:bbar}. This makes the distinction clear: $\bar a$ is local, while $\bar b$ depends on the outer integration prescription.

\section{Angular units and observational inputs}\label{app:observational}

The gravitational angular scale is
\begin{equation}
 \theta_g[\mu\mathrm{as}]
 =\frac{GM}{c^2D}\,(206265\times10^6).
\end{equation}
The numerical calculation uses
\begin{equation}
 G=6.67430\times10^{-11}\,\mathrm{m^3\,kg^{-1}\,s^{-2}},
 \quad c=2.99792458\times10^8\,\mathrm{m\,s^{-1}},
\end{equation}
\begin{equation}
 M_\odot=1.98847\times10^{30}\,\mathrm{kg},
 \quad 1\,\mathrm{pc}=3.08567758149\times10^{16}\,\mathrm{m}.
\end{equation}
For the benchmark plots we adopt $M=6.5\times10^9M_\odot$, $D=16.8$ Mpc for M87*, and $M=4.0\times10^6M_\odot$, $D=7.97$ kpc for Sgr~A*. These central values are used only to convert $b_c/M$ into microarcseconds. The observational intervals come from the EHT analyses cited in Sec.~\ref{sec:lens}. Equation~\eqref{eq:errorprop} should be used when a specific correlated mass-distance prior is adopted.

\section{Numerical procedure and validation}\label{app:numerics}

We work in units of $M$ and evaluate the benchmark path directly with dimensionless variables. We locate all sign changes of the horizon and photon-sphere equations on logarithmic radial grids and refine them with Brent's method. A photon-sphere candidate is accepted only if it lies in the chosen static region and satisfies the instability condition in Eq.~\eqref{eq:unstable}.

For the weak-lensing check, we use the exact geodesic of the locally matched magnetic-PFDM metric. The turning point is found from
\begin{equation}
 1-\frac{b^2B(r_0)}{r_0^2}=0.
\end{equation}
The change of variable $r=r_0+t^2$ removes the square-root singularity at the endpoint. Table~\ref{tab:weakvalidation} compares the exact angle with Eq.~\eqref{eq:alphaweak}.

\begin{table}[t]
 \centering
 \caption{Weak-field validation for $\eta=1$. Angles are in radians.}
 \label{tab:weakvalidation}
 \begin{tabular}{c c c c}
\toprule
$b/M$ & $\alpha_{\rm exact}$ & $\alpha_{\rm weak}$ & relative difference\\
\midrule
60 &  0.08297194 &  0.08143399 &  $1.854\times 10^{-2}$\\
100 & 0.04912584 & 0.04858585 & $1.099\times 10^{-2}$\\
160 & 0.03059701 & 0.03038379 & $6.968\times 10^{-3}$\\
230 & 0.02130233 & 0.02119690 & $4.949\times 10^{-3}$\\
290 & 0.01692222 & 0.01685473 & $3.989\times 10^{-3}$\\
\bottomrule
\end{tabular}
\end{table}

For strong lensing, we use the same endpoint substitution in the exact radial integral. The coefficient $\bar a$ is calculated directly from Eq.~\eqref{eq:abar}. To check $\bar b$, we sample $2\times10^{-6}\leq b/b_c-1\leq10^{-4}$ and take the median of
\begin{equation}
 \alpha(b)+\bar a\ln\!\left(\frac{b}{b_c}-1\right).
\end{equation}
We then reconstruct the regular part with Eq.~\eqref{eq:bbar}. Table~\ref{tab:strongconvergence} shows how the result changes with the matching radius.

\begin{table}[t]
 \centering
 \caption{Convergence of the regular strong-deflection quantities with the matching radius.}
 \label{tab:strongconvergence}
 \begin{tabular}{c c c c}
\toprule
$\eta$ & $r_{\rm max}/M$ & $\bar b$ & $I_R$\\
\midrule
0.0 & 1000 & $-0.410558$ & 0.939275\\
0.0 & 5000 & $-0.402244$ & 0.947589\\
0.0 & 20000 & $-0.400686$ & 0.949148\\
1.0 & 1000 & $-0.400064$ & 0.919366\\
1.0 & 5000 & $-0.390839$ & 0.928591\\
1.0 & 20000 & $-0.389109$ & 0.930321\\
\bottomrule
\end{tabular}
\end{table}

\bibliography{refs}

\end{document}